\documentclass[twocolumn]{aastex61}
\usepackage{multirow}
\usepackage{booktabs}
\usepackage{array}
\def\apjl{ApJL}% % Astrophysical Journal, Letters 
\def\apjs{ApJS}% % Astrophysical Journal, Supplement 
\def\ao{Appl.~Opt.}% % Applied Optics 
\def\aap{A\&A}% % Astronomy and Astrophysics 
\DeclareMathAlphabet{\mathsc}{OT1}{cmr}{m}{sc}
\def\testbx{bx}%
\DeclareRobustCommand{\ion}[2]{%
\relax\ifmmode
\ifx\testbx\f@series
{\mathbf{#1\,\mathsc{#2}}}\else
{\mathrm{#1\,\mathsc{#2}}}\fi
\else\textup{#1\,{\mdseries\textsc{#2}}}%
\fi}
\newcommand{\Hi} {\ion{H}{i}}

\newcommand{\ha} {\mbox{H$\alpha$}}
\newcommand{\hb} {\mbox{H$\beta$}}

\newcommand{\Hei} {\ion{He}{i}}

\newcommand{\Oi} {\ion{O}{i}}

\newcommand{\Nai} {\ion{Na}{i}}

\newcommand{\Siii} {\ion{Si}{ii}}

\newcommand{\Caii} {\ion{Ca}{ii}}
\newcommand{\Scii} {\ion{Sc}{ii}}
\newcommand{\Tiii} {\ion{Ti}{ii}}

\newcommand{\Feii} {\ion{Fe}{ii}}

\newcommand{\Baii} {\ion{Ba}{ii}}

\newcommand{\sn}{ASASSN-15nx}
\newcommand{\host}{GALEXASC~J044353.08-094205.8}
\newcommand{\EpEpoch}{JD~2457219.10}

\newcommand{\synow}{\textsc{synow}}

\newcommand{\iraf}{\texttt{IRAF}}
\newcommand{\daophot}{\textsc{daophot}}

\newcommand{\kpc}{\ensuremath{\mbox{~kpc}}}
\newcommand{\ebv}{\mbox{$E(B-V)$}}

\newcommand{\maghundred}{\mbox{mag (100 d)$ ^{-1} $}}
\newcommand{\msun}{\mbox{M$_{\odot}$}}

\newcommand{\kms}{\mbox{$\rm{\,km\,s^{-1}}$}}
\newcommand{\kkms}{\mbox{$\times10^3\rm{\,km\,s^{-1}}$}}

\newcommand{\nickel}{\mbox{$^{56}$Ni}}
\newcommand{\cobalt}{\mbox{$^{56}$Co}}
\newcommand{\iron}{\mbox{$^{56}$Fe}}

\newcommand{\ld}{\mbox{$\lambda$}}
\newcommand{\ldld}{\mbox{$\lambda\lambda$}}
 
\begin{document}

\title{\sn: A luminous Type II supernova with a ``perfect" linear decline} 
\author{Subhash Bose} \affil{Kavli Institute for Astronomy and Astrophysics, Peking University, Yi He Yuan Road 5, Hai Dian District, Beijing 100871, China.}
\author{Subo Dong} \affil{Kavli Institute for Astronomy and Astrophysics, Peking University, Yi He Yuan Road 5, Hai Dian District, Beijing 100871, China.}
\author{C. S. Kochanek} \affil{Department of Astronomy, The Ohio State University, 140 W. 18th Avenue, Columbus, OH 43210, USA.} \affil{Center for Cosmology and AstroParticle Physics (CCAPP), The Ohio State University, 191 W. Woodruff Avenue, Columbus, OH 43210, USA.}
\author{Andrea Pastorello} \affil{INAF-Osservatorio Astronomico di Padova, Vicolo dell'Osservatorio 5, I-35122 Padova, Italy}
\author{Boaz Katz} \affil{Weizmann Institute of Science, Rehovot, Israel}
\author{David Bersier} \affil{Astrophysics Research Institute, Liverpool Science Park, 146 Brownlow Hill, Liverpool L3 5RF, UK 0000-0001-7485-3020}
\author{Jennifer E. Andrews} \affil{Steward Observatory, University of Arizona, Tucson, AZ 85721, USA}
\author{J. L. Prieto} \affil{N\'ucleo de Astronom\'ia de la Facultad de Ingenier\'ia y Ciencias, Universidad Diego Portales, Av. Ej \'ercito 441, Santiago, Chile} \affil{Millennium Institute of Astrophysics, Santiago, Chile.}
\author{K. Z. Stanek} \affil{Department of Astronomy, The Ohio State University, 140 W. 18th Avenue, Columbus, OH 43210, USA.} \affil{Center for Cosmology and AstroParticle Physics (CCAPP), The Ohio State University, 191 W. Woodruff Avenue, Columbus, OH 43210, USA.}
\author{B. J. Shappee} \affil{Institute for Astronomy, University of Hawaii, 2680 Woodlawn Drive, Honolulu, HI 96822, USA}
\author{Nathan Smith} \affil{Steward Observatory, University of Arizona, Tucson, AZ 85721, USA}
\author{Juna Kollmeier} \affil{Carnegie Observatories, 813 Santa Barbara Street, Pasadena, CA 91101, USA}
\author{Stefano Benetti} \affil{INAF-Osservatorio Astronomico di Padova, Vicolo dell'Osservatorio 5, I-35122 Padova, Italy}
\author{E. Cappellaro} \affil{INAF-Osservatorio Astronomico di Padova, Vicolo dell'Osservatorio 5, I-35122 Padova, Italy}
\author{Ping Chen} \affil{Kavli Institute for Astronomy and Astrophysics, Peking University, Yi He Yuan Road 5, Hai Dian District, Beijing 100871, China.}
\author{N. Elias-Rosa} \affil{INAF-Osservatorio Astronomico di Padova, Vicolo dell'Osservatorio 5, I-35122 Padova, Italy}
\author{Peter Milne} \affil{Steward Observatory, University of Arizona, Tucson, AZ 85721, USA}
\author{Antonia Morales-Garoffolo} \affil{Department of Applied Physics, University of C\'adiz, Campus of Puerto Real, E-11510 C\'adiz, Spain)}
\author{Leonardo Tartaglia} \affil{Department of Astronomy and The Oskar Klein Centre, AlbaNova University 
	Center, Stockholm University, SE-106 91 Stockholm, Sweden}
\author{L. Tomasella} \affil{INAF-Osservatorio Astronomico di Padova, Vicolo dell'Osservatorio 5, I-35122 Padova, Italy}
\author{Christopher Bilinski} \affil{Steward Observatory, University of Arizona, Tucson, AZ 85721, USA}
\author{Joseph Brimacombe} \affil{Coral Towers Observatory, Cairns, Queensland 4870, Australia}
\author{Stephan Frank} \affil{Department of Astronomy, The Ohio State University, 140 W. 18th Avenue, Columbus, OH 43210, USA.}
\author{T. W.-S. Holoien} \affil{Carnegie Observatories, 813 Santa Barbara Street, Pasadena, CA 91101, USA}
\author{Charles D. Kilpatrick} \affil{Department of Astronomy and Astrophysics, University of California, Santa Cruz, CA 95064, USA}
\author{Seiichiro Kiyota} \affil{Variable Stars Observers League in Japan (VSOLJ), 7-1 Kitahatsutomi, Kamagaya 273-0126, Japan}
\author{Barry F. Madore} \affil{Carnegie Observatories, 813 Santa Barbara Street, Pasadena, CA 91101, USA}
\author{Jeffrey A. Rich} \affil{Carnegie Observatories, 813 Santa Barbara Street, Pasadena, CA 91101, USA}

\correspondingauthor{Subo Dong, Subhash Bose}
\email{dongsubo@pku.edu.cn, email@subhashbose.com}

\begin{abstract}
We report a luminous Type II supernova, \sn, with a peak luminosity of $ M_V=-20 $ mag, that is between typical core-collapse supernovae and super-luminous supernovae. The post-peak optical light curves show a long, linear decline with a steep slope of 2.5 \maghundred (i.e., an exponential decline in flux), through the end of observations at phase $\approx 260\,{\rm d}$. In contrast, the light curves of hydrogen rich supernovae (SNe~II-P/L) 
always show breaks in their light curves at phase $\sim 100\,{\rm d}$, before settling onto $^{56}$Co radioactive decay tails with a decline rate of about 1 \maghundred. The spectra of \sn\, do not exhibit the narrow emission-line features characteristic of Type IIn SNe, which can have a wide variety of light-curve shapes usually attributed to strong interactions with a dense circumstellar medium (CSM). \sn\ has a number of spectroscopic peculiarities, including a relatively weak and triangularly-shaped \ha\ emission profile with no absorption component. 
The physical origin of these peculiarities is unclear, but the long and linear post-peak light curve without a break suggests a single dominant powering mechanism. Decay of a large amount of \nickel\ ($ M_{Ni}=1.6\pm0.2 $\msun) can power the light curve of \sn, and the steep light-curve slope requires substantial $\gamma$-ray escape from the ejecta, which is possible given a low-mass hydrogen envelope for the progenitor.  Another possibility is strong CSM interactions powering the light curve, but the CSM needs to be sculpted to produce the unique light-curve shape and to avoid producing SN IIn-like narrow emission lines. 
\end{abstract}


% sec:intro
%______________________________________________________________________________________________

\section{Introduction} \label{sec:intro}

Core-collapse supernovae (CCSNe) are generally believed to originate from the collapse of massive stars with zero age main sequence (ZAMS) masses $ M_{\rm ZAMS} \gtrsim 8 \msun $. The properties of the resulting transient depend strongly on the mass and composition of the star at death. In particular, 
Type II supernovae (SNe) represent the broad subclass of CCSNe which have retained a substantial amount of hydrogen envelope at the time of explosion. Their spectra show characteristic prominent hydrogen Balmer lines with P-Cygni profiles, while the other subclasses of CCSNe (Ib and Ic) are characterized by the absence of hydrogen in their spectra.

Traditionally, these hydrogen-rich SNe are classified into two major subclasses, Type II-P and II-L \citep{1979A&A....72..287B, 1997ARA&A..35..309F}, based on their light curve shapes in the photospheric phase. In this classification scheme, the light curve of a SN II-P has a plateau with almost constant brightness for a period of nearly $100$ days, whereas the light curve of a SN II-L declines  linearly in  magnitude after its peak. Various attempts have been made to refine the classifying criteria of these two subclass \citep[see, e.g.,][]{2012ApJ...756L..30A, 2014MNRAS.445..554F}. However, with more detections of SNe-II, it has been realized that SN II light curves are too diverse to perfectly divide into ``plateau'' or ``linear'' shapes \citep[e.g][]{2016MNRAS.459.3939V,2016MNRAS.455.2712B,ASASSN13co}, and that the distribution of light-curve shapes may be continuous rather than bimodal \citep{2014ApJ...786...67A}.
A continuous distribution of light curve decline rates suggests a continuum in the ejecta parameters controlling the light-curve shape (e.g., the progenitor density profile according to \citealt{ET}). Hereafter, we would refer them in an unified subclass as Type II-P/L SNe.

At the end of the photospheric phase, there is a sudden change in the light-curve shape of SN II-P/L to an exponential tail with a decline rate typically of about 0.98 \maghundred, which is the rate of energy deposition from the radioactive decay chain of $\nickel\rightarrow\cobalt\rightarrow\iron$. Excluding SNe with strong late-time CSM interactions, all common supernovae (CCSNe and SNe Ia) light curves show this nuclear-decay dominated phase at late times. 
The rate of energy deposition into the ejecta is governed by the nuclear decay rate, while the light curve shape can be modified based on how the ejecta traps and thermalizes the $ \gamma $-ray photons released from the nuclear decay.
For SNe II-P/L, the ejecta mass is typically large enough to efficiently absorb and thermalize the decay energy, leading to a luminosity decline rate close to the nuclear decay rate. With the end of the recombination dominated photospheric phase and the onset of the radioactive tail phase, there is a sharp transition in the optical light curve. For all SN II-P/L with well-covered late-time light curves, this transition phase is always seen \citep{2014ApJ...786...67A}.

CCSNe originate from a wide range of progenitors, and their observed properties also have a large diversity. The peak absolute magnitude of Type Ib/c and Type II-P/L SNe typically lie within a broad range of $M_V\sim -14 $ to $ -18.5 $ mag \citep{lisample}. In the last decade we have also seen the emergence of a possibly new class of events, super-luminous  supernovae \citep[SLSNe; e.g.][]{2007ApJ...668L..99Q,2007ApJ...666.1116S,2016Sci...351..257D,2018ApJ...853...57B}. They are ten to hundred times more luminous than typical CCSNe and peak at $ M_V < -21 $ mag. Their explosion physics and powering mechanisms are not yet understood though many hypothesize that their progenitors may be stars more massive than those of common CCSNe \citep{2012Sci...337..927G}. It is an open question as to whether there is gap in the SN luminosity function between those of common SNe and the SLSNe \citep{2016ApJ...819...35A}, and the answer to this question may indicate whether the progenitor masses of SLSNe are just an extension of the normal SNe. Only a few SNe have been discovered with intermediate luminosity (e.g., PTF10iam with $ M_{V, \rm peak}\approx-20$ mag; \citealt{2016ApJ...819...35A}). 

Here we report the latest addition to the rare group of events with luminosities in between that of typical CCSNe and SLSNe, \sn. We present the discovery and follow-up observations of this Type II SN to late time, and we find that unlike any known SN II-P/L, its late-time light curves do not show the transition to a nuclear decay tail.

\begin{figure}
\centering
\includegraphics[width=7cm]{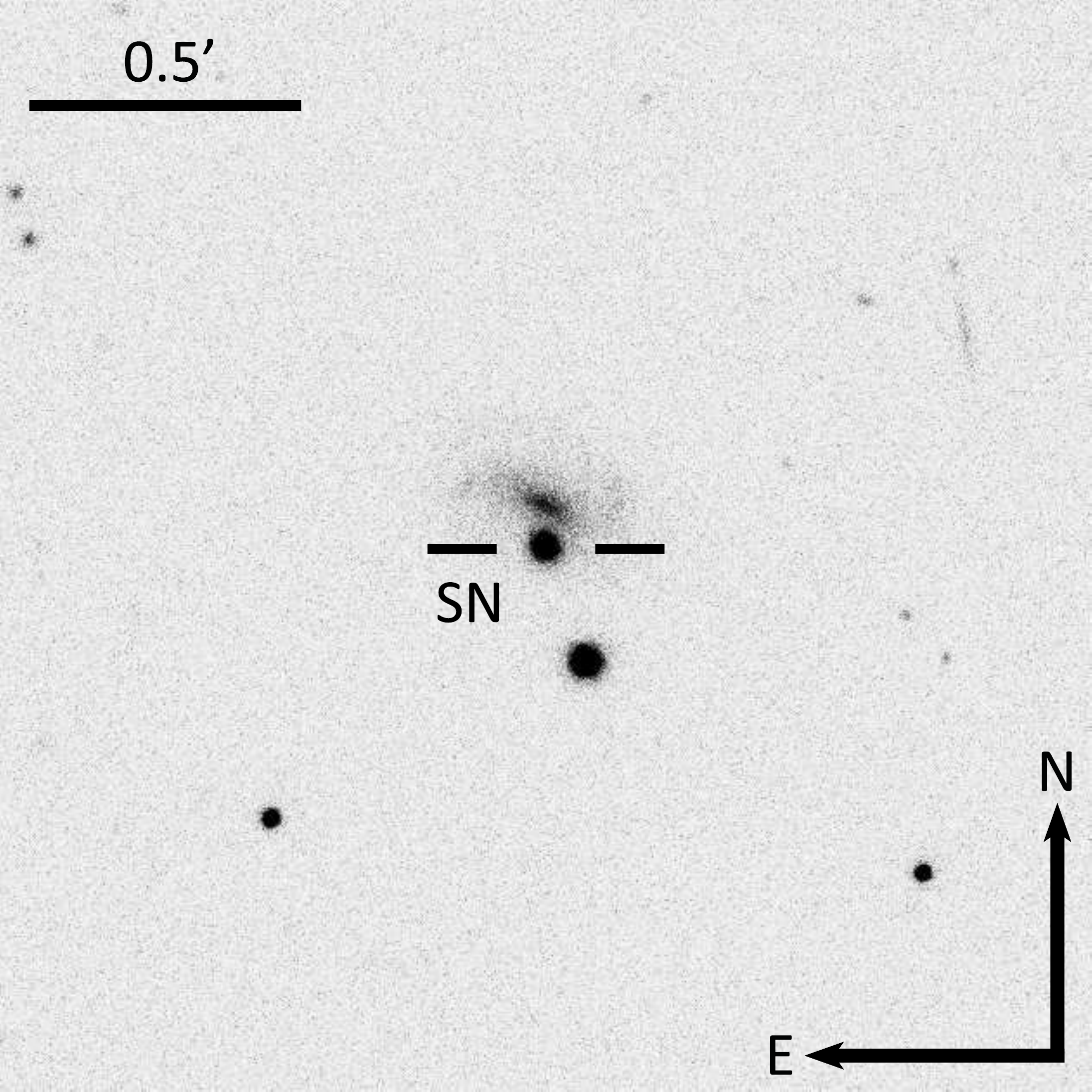}
\caption{A 2\arcmin$\times$2\arcmin\ $g'$-band image from the 2.6m Nordic Optical Telescope showing \sn\ and its host \host. } 
\label{fig:snid}
\end{figure}

\sn\ was discovered \citep{2015ATel.7895....1K,2017MNRAS.467.1098H} in the galaxy \host\ (see Fig.~\ref{fig:snid} for an image of the supernova and its host galaxy) on August 8, 2016 during the ongoing All-Sky Automated Survey for SuperNovae \citep[ASASSN;][]{2014ApJ...788...48S}, using the quadruple 14-cm ``Brutus" telescope at the LCO facility on Haleakala, Hawaii. 
The ASAS-SN survey regularly scans the entire visible sky for bright supernovae and other extragalactic transients down to $V\sim 17\,$mag, and the ASAS-SN discoveries are minimally biased by host galaxy properties \citep{2017MNRAS.467.1098H}.
Nearly 100\% of the ASAS-SN supernovae also have spectroscopic classifications \citep{asassn201314, 2017MNRAS.467.1098H, 2017MNRAS.471.4966H}. As a result the ASAS-SN survey provides an unprecedented, spectroscopically complete and host-unbiased sample from an un-targeted survey to study supernova statistics. It has also found a range of unusual transients that likely would have been missed in  many other surveys \citep[e.g.,][]{2016Sci...351..257D, 2016MNRAS.455.2918H, 2018ApJ...853...57B}.
\sn\ is located at $\alpha = 04^{\rm h} 43^{\rm m} 53\fs19$ $\delta = -09\degr 42\arcmin 11\farcs22$, which is offset by 0\farcs9 E and 5\farcs1 S of the center of the $z=0.02823$ (see \S\ref{sec:ext}; $ D_L= 127.5$ Mpc) host galaxy \host\ (or PGC 987599; $\alpha = 04^{\rm h} 43^{\rm m} 53\fs13$, $\delta = -09\degr 42\arcmin 06\farcs1$). The offset from the host center is 3.2~\kpc.

The first detection of \sn\ was on July 16.42, 2015 UTC (JD 2457219.92). We adopt July 15.60, 2015 (\EpEpoch$ \pm 2.00 $) as the explosion epoch based on fitting the early rising of the light curve with the analytical model from \citet{2011ApJ...728...63R}. We used this as the reference epoch throughout the paper (see \S\ref{sec:ext} for further details on the method of constrain the explosion epoch). The object was classified by \cite{2015ATel.8016....1E} as a Type II supernova based on the presence of an \ha\ emission line, and it was noted that the \ha\ emission had a peculiar, triangular profile and metal lines that appeared to be unusually strong. 

We provide brief descriptions of the data collected for \sn\ in \S\ref{sec:data}. We discuss how we estimate explosion epoch, distance and extinction in \S\ref{sec:ext}. In Sections \S\ref{sec:lc} and \S\ref{sec:sp}, we perform detailed photometric and spectroscopic characterization of the SN and also identify various peculiarities, which are summarized in \S\ref{sec:sum}. Finally. in \S\ref{sec:discuss}, we discuss various scenarios that may explain the unique properties of \sn.

%\input{./host.tex}

% sec:obs
% __________________________________________________________________________

\section{Data} \label{sec:data}

We initiated multi-band photometric and spectroscopic observations soon after the discovery (+25d) and continued the observations of \sn\ until +262d. Photometric data were obtained from the ASAS-SN quadruple 14-cm "Brutus" telescope, the Las Cumbres Observatory 1.0m telescope network \citep{2013PASP..125.1031B}, the 1.8m Copernico, 0.8m TJO, 2.4m MDM, 2.6m NOT, 2.0m Liverpool telescope, 6.5m Magellan Baade and 0.6m Super-Lotis telescopes. The data were obtained in the Johnson-Cousins $BVRI$ and SDSS $gri$ broadband filters. The images were reduced using standard  \iraf\ tasks, and PSF photometry was performed using the \daophot\ package. The PSF radius and background extraction region were adjusted according to the FWHM of the image. Photometric calibrations are done using APASS \citep[DR9;][]{2016yCat.2336....0H} standards available in the field of observation. The $R$ and $I$-band standards were converted from Sloan $gri$ magnitudes using the transformation relation given by \cite{2005AAS...20713308L}. The photometric data of \sn\ are reported in Table~\ref{tab:photsn_simple}.

Medium to low resolution spectroscopic observations were made using AFOSC mounted on the 1.8m Copernico, the Boller \& Chivens Spectrograph on the 2.3m Bok, ALFOSC on the 2.6m NOT, the Blue Channel spectrograph on the 6.5m MMT, IMACS and MagE on the 6.5m Magellan Baade telescope, the Boller \& Chivens Spectrograph on the 2.5m Irenee du Pont and MODS on the LBT with an effective diameter of 11.9m. All observations were performed in long-slit mode and spectroscopic reductions were done using standard \iraf\ tasks. The medium-resolution spectra from MODS were reduced using the \textrm{modsIDL} pipeline. The LBT MODS observation on Jan 2.20, 2016 was also used to estimate the redshift of the host galaxy (see \S\ref{sec:ext}), with the slit aligned to cross both the host nucleus and the SN. The spectroscopic observations are summarized in Table~\ref{tab:speclog}.

%_________________________________________________________________________________
\section{Explosion epoch, distance and extinction} \label{sec:ext}
\begin{figure}
 	\centering
 	\includegraphics[width=0.8\linewidth]{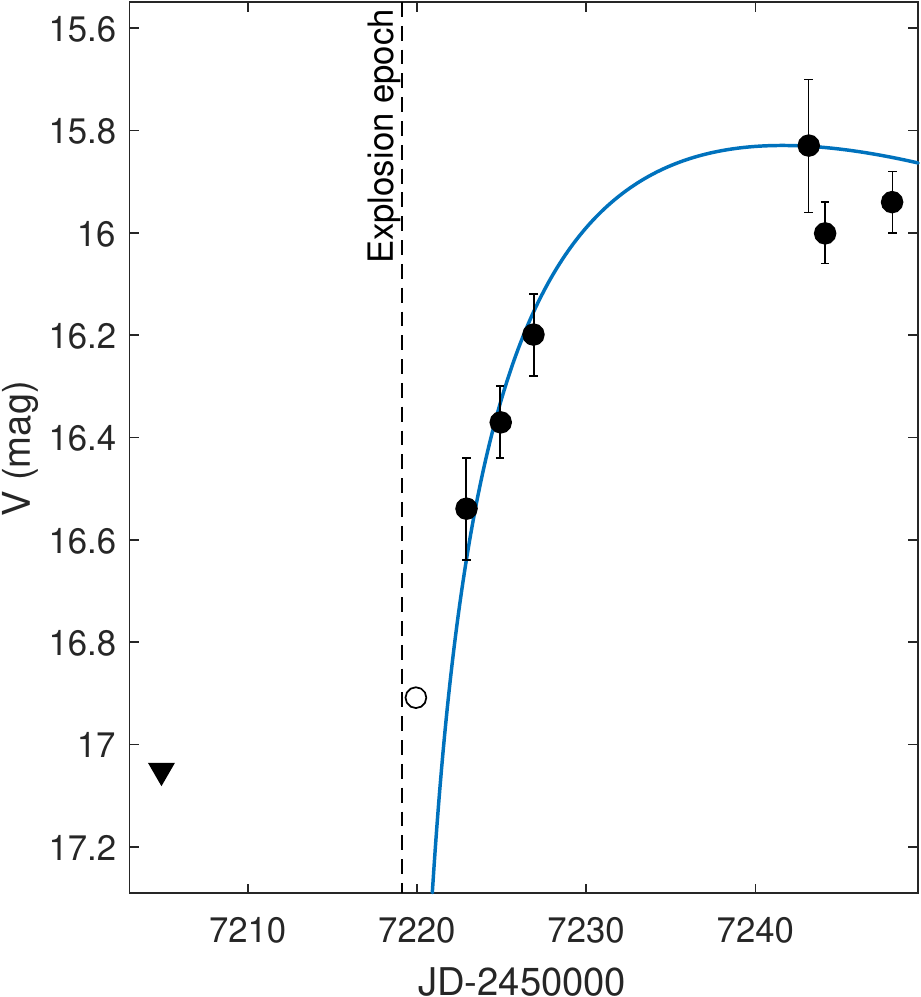}\\
 	\caption{ Modeling of the early \textit{V}-band light light curve of \sn\ to constrain the explosion epoch. The filled triangle shows the last ASAS-SN non-detection.  The open circle represent the first confirmed detection of the SN, but with uncertain photometry. The estimated explosion epoch 0.8 days prior to the first detection and is indicated by the dashed line.}
 	\label{fig:lc_rise_model}
 \end{figure}

The first confirmed detection of \sn\ was on JD~2457219.92 (July 16.42, 2015 UTC), and the last non-detection was about 15 days earlier on JD~2457204.94 with a limiting magnitude of \textit{V}$=$17.1 mag. This 15-day gap prevents us from rigorously constraining the explosion epoch directly from observations. However, subsequent observations after the initial detection captured the early rise of the light curve. We modeled these data following \citet{2011ApJ...728...63R}, which is strictly applicable within a couple of weeks after explosion. First we construct the black-body SED using the temperature and radius from the \cite{2011ApJ...728...63R} prescription for a red-supergiant progenitor. The SED is then redshifted and corrected for extinction. The model SED evolution can be represented as

\[ F(\lambda,t)=\frac{A\cdot (t-t_0)^{1.62}}{\lambda^5\, \left[{\rm exp}\left({\frac{B\cdot (t-t_0)^{0.45}}{\lambda}}\right) - 1\right]}, \]
\\
where $ t_0 $ is the explosion epoch, A and B are free parameters, and $ \lambda $ and $ t $ have the usual meanings of wavelength and time. The resulting SED evolution is then convolved with \textit{V}-band filter response to obtain the model light curve. The fit to the early \textit{V}-band observations is shown in Fig.~\ref{fig:lc_rise_model}. Even though the nominal detection significance of the first detection is high, the quality of the image is poor and we exclude it from the model fitting. 
The data for the subsequent three epochs are cleaner and provide most of the model constraints leading to an
estimated explosion epoch of \EpEpoch$\pm2.00 $ (July 15.60, 2015 UTC), which is $ \sim0.8 $ days prior to our first detection. We adopt this as the reference epoch throughout the paper. The estimated rise time from explosion to \textit{V}-band peak is then $ \approx22 $ days.

 % fig:lc_rise_model 
 %____________________________________________________________________________

\begin{figure}
 \centering
  \includegraphics[width=\linewidth]{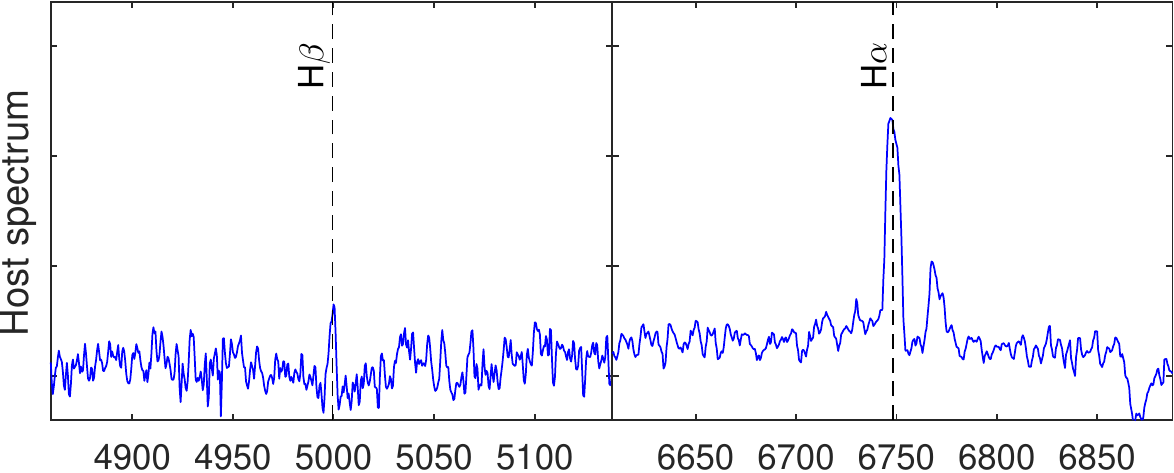}\\
%  \vspace{0.2cm}
%  \includegraphics[width=\linewidth]{./plot_naid.pdf}%
 \caption{Spectrum for the host nucleus with narrow \ha\ and \hb\ lines at a redshift $ z=0.02823 $. 
 %The bottom panel shows the marginal detection of the blend of \Nai~D absorption lines due to Milky-Way, whereas at the redshifted wavelength for the host no significant feature is present.
 }
 \label{fig:emission.naid}
 \end{figure}

The host galaxy of \sn\ does not have any archival redshift or distance estimate. We took a medium-resolution LBT/MODS spectrum with the slit crossing the nucleus of the host galaxy on 165.9d. The spectrum revealed narrow \Hi\ lines (see Fig.~\ref{fig:emission.naid}) with \ha\  being the most prominent feature at 6748.1\AA, corresponding to a host redshift of $ z=0.02823 $. The weaker \hb\ line i also detected at 4999.4\AA\ corresponding to a $ z=0.02840 $. These two values are consistent and we adopt $ z=0.02823 $ from the strong \ha\ line. This redshift is also consistent with that inferred from the faint, narrow \ha\ emission visible on top of the broad \ha\ P-Cygni profile in four late SN spectra taken between 166d and 262d. The corresponding luminosity distance and distance modulus are $D_L= 127.5\pm1.7 $ Mpc and $ \rm{DM}=35.53\pm0.03 $ mag, assuming a flat cosmology with $ H_0=67.7~\rm{km~s^{-1}Mpc^{-1}} $ and $ \Omega_m=0.308 $ \citep{2016A&A...594A..13P}.

In the 165.9d SN spectrum, we detect Galactic \Nai~D absorption at 5893\AA. We do not detect any \Nai~D absorption feature at the redshift of the host, which indicates that the host extinction is likely negligible.
Therefore we adopt a total line-of-sight reddening of $ \ebv=0.07$ mag \citep{2011ApJ...737..103S} entirely due to the Milky Way, which translates into $ A_V=0.22$ mag assuming $ R_V=3.1 $ \citep{1989ApJ...345..245C}.

 % fig:naid.ech 
 %____________________________________________________________________________

% sec:lc
%_________________________________________________________________________________
\section{Light curve} \label{sec:lc} %completed

 \subsection{Light curve evolution and comparison} \label{sec:lc.app}

\begin{figure}
	\centering
	%\resizebox{10cm}{10cm}{\includegraphics{lightcurve.ps}}%
	\includegraphics[width=\linewidth]{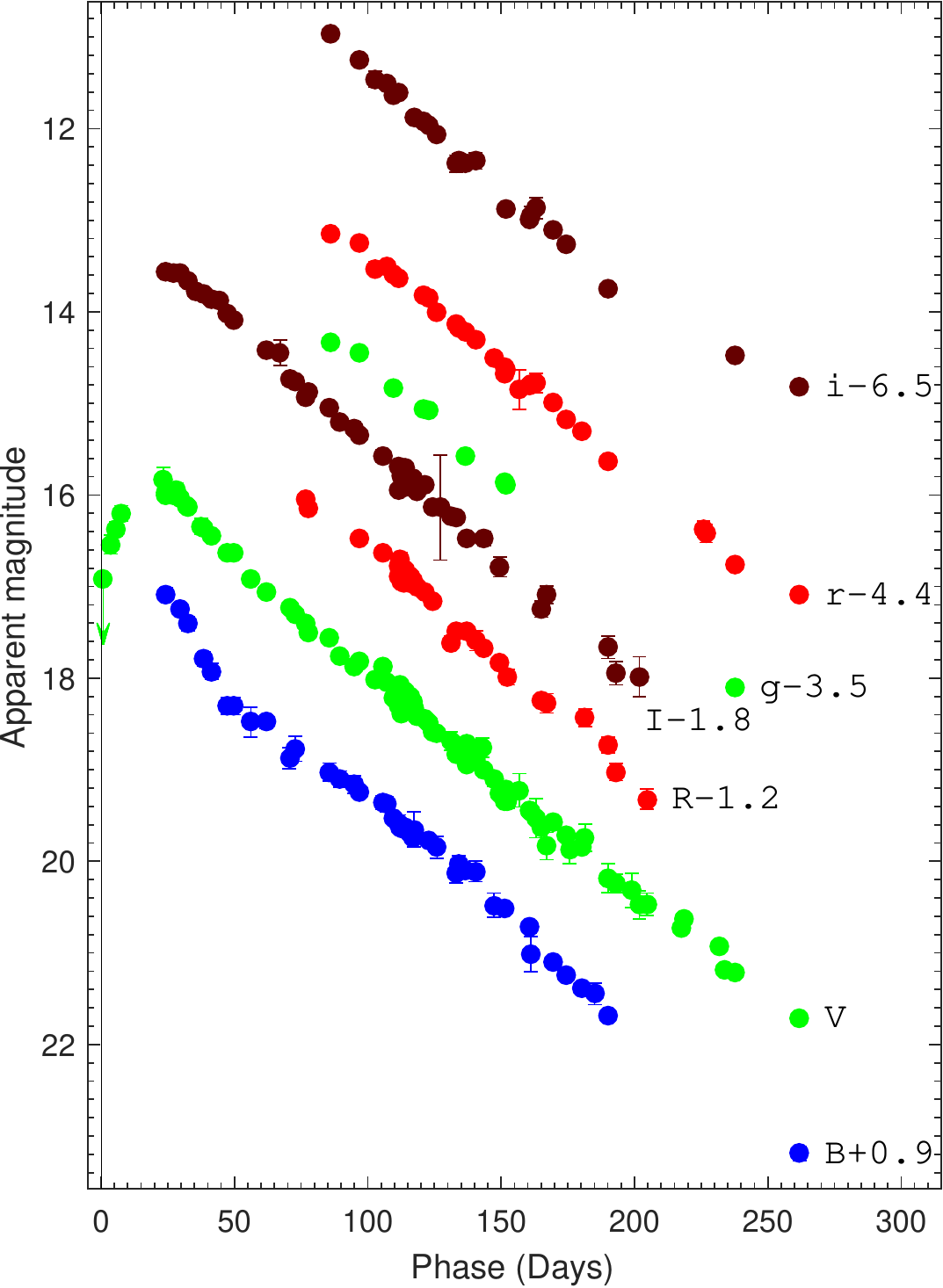}
	\caption{The photometric light curves in the Johnson-Cousins \textit{BVRI} and SDSS \textit{gri} bands. The light curves are vertically shifted for clarity.}
	\label{fig:lc.app}
\end{figure}

The most unique feature of \sn\ is its long-lasting, fast-declining linear light curve during the entire phase of evolution following the maximum as shown in Fig.~\ref{fig:lc.app}. Post-maximum linear decline at a constant rate is observed in all photometric bands with only the exception of the \textit{B}-band, which has a steeper slope for $ \lesssim50 $d. This exceptional long and nearly ``perfectly" continuous linear decline in most optical bands has not been seen in any other SNe  observed to date. The rest-frame light curve decline rates are $ 2.48\pm 0.03$, $ 2.53\pm 0.08 $, $ 2.65\pm0.04 $, $ 2.82\pm0.25 $, $ 2.47\pm0.10 $, $ 2.51\pm0.09 $ \maghundred\, in the \textit{V, R, I, g, r}, and \textit{i} bands respectively. The \textit{B} band light curve slope is $ 5.28\pm0.28 $ \maghundred\ for $ <52 $d and $ 2.46\pm0.07 $ \maghundred\ afterwards.

% fig:lc.app
%_________________________________________________________________

\begin{figure}
	\centering
	%\hspace{-0.5cm}
	\includegraphics[width=1.02\linewidth]{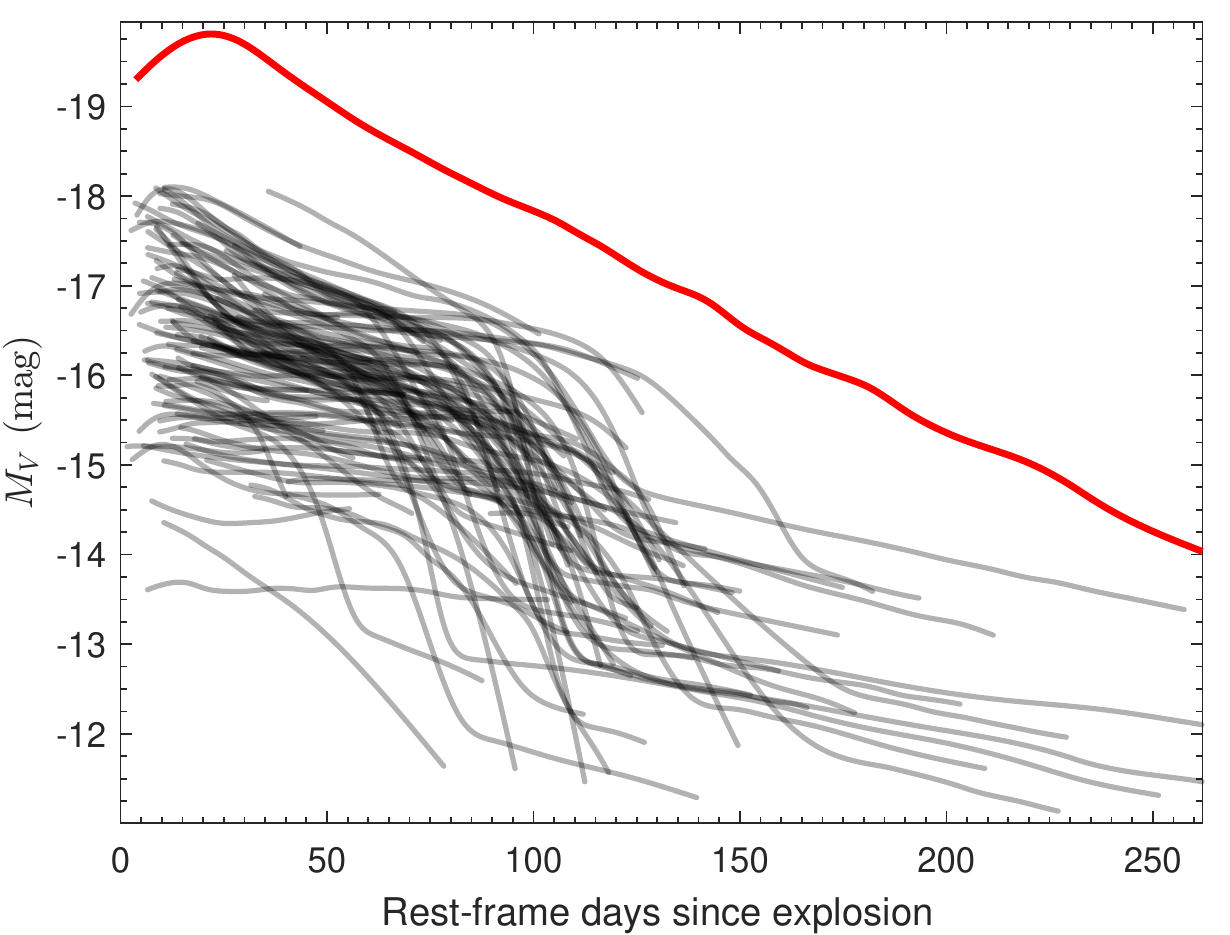}
	\caption{The absolute \textit{V}-band light curve of \sn\ (red) is compared with the sample of Type II SNe presented by \cite{2014ApJ...786...67A}.}
	\label{fig:mv.bulk}
\end{figure}

We compare the absolute \textit{V}-band ($M_V$) light curve of \sn\ with those for 116 Type II-P/L SNe from \cite{2014ApJ...786...67A} in Fig.~\ref{fig:mv.bulk}. The comparison shows that the SN clearly stands out from the sample in terms of both absolute magnitude and the nearly perfect, long linear decline of the light curve. The \textit{V}-band maximum absolute magnitude observed for \sn\ is $ -19.92\pm0.06 $~mag making it $ \sim2.8 $ mag  brighter at $ +50 $d than typical Type II-P/L SNe in the sample. 

%todo: convert to eps

\begin{figure*}
	\centering
	%\resizebox{10cm}{10cm}{\includegraphics{abs_lightcurve.ps}}%
	\includegraphics[width=16cm]{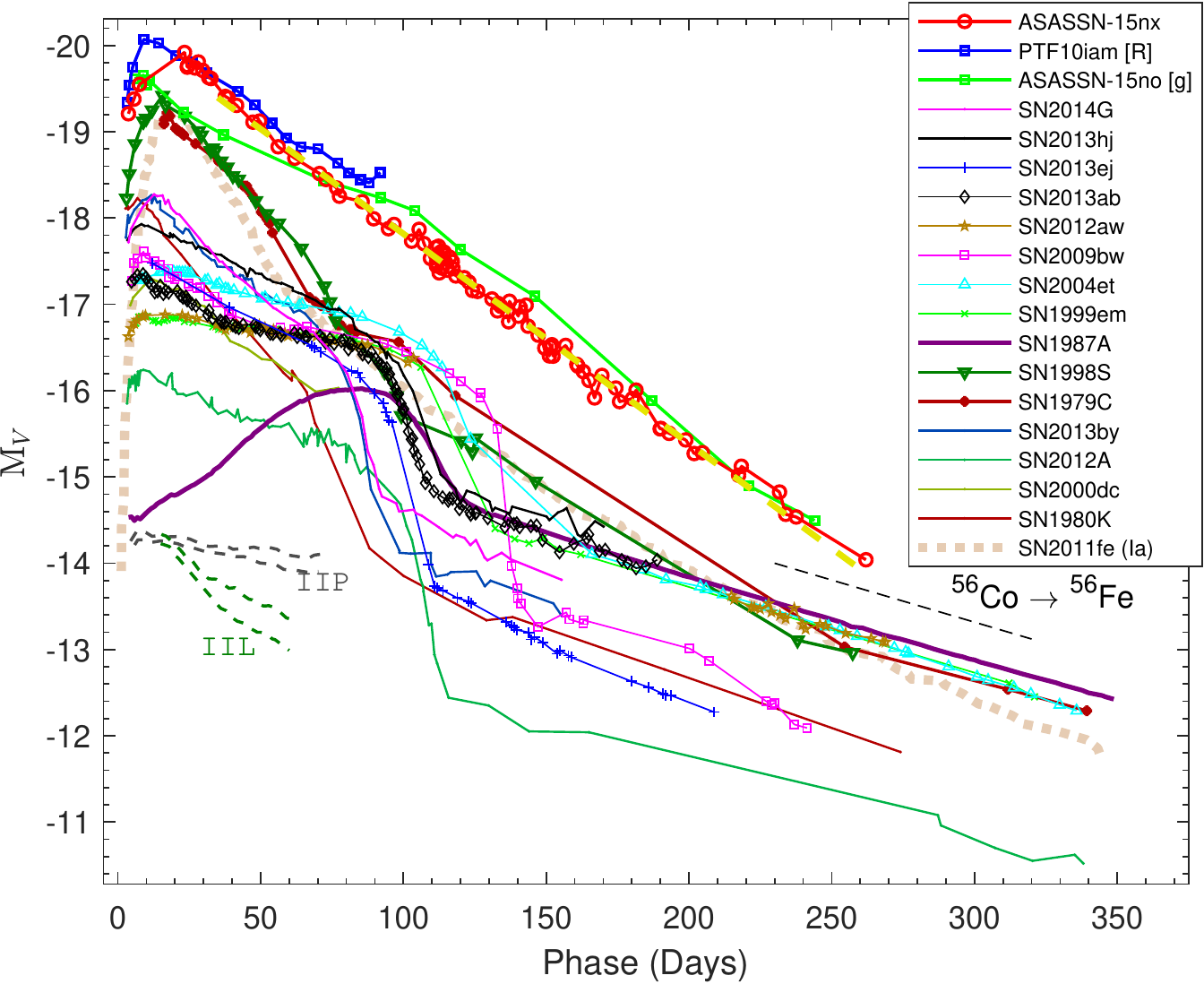}%
	\caption{
		The absolute $ V $-band light curve of \sn\ as compared to other Type II SNe and the Type Ia SN 2011fe.
		%Comparison of M$_{V}$ light curve of {\sn} with other Type II-P SNe. 
		The exponential decline of a light curve following the radioactive decay law for \cobalt$ \rightarrow $\iron\ is shown with a black dashed line. A best fit straight line (yellow dashed) is shown on top of \sn\ light curve to emphasize the linearity of decline. On the bottom left side, pairs of dashed lines in each gray and green colors represent the slope range for the Type II-P and II-L SNe templates as given by \citet{2014MNRAS.445..554F}. The adopted explosion time in JD-2400000, 
		distance in Mpc, \ebv\ in mag and 
		the reference for observed \textit{V}-band magnitude, respectively, are : 
		SN 1979C  -- 43970.5, 16.0, 0.31;  \citet{1982A&A...116...43B,1981PASP...93...36D}; 
		SN 1980K  -- 44540.5, 5.5, 0.30;  \citet{1982A&A...116...35B}, NED database; 
		SN 1987A  -- 46849.8, 0.05, 0.16;  \citet{1990AJ.....99.1146H}; 
		SN 1999em -- 51475.6, 11.7, 0.10; \citet{2002PASP..114...35L,2003MNRAS.338..939E};
		SN 2000dc -- 51762.4, 49.0, 0.07; \citet{2014MNRAS.445..554F}, NED database;
		SN 2004et -- 53270.5, 5.4, 0.41; \citet{2006MNRAS.372.1315S}; 
		SN 2009bw -- 54916.5, 20.2, 0.31; \citet{2012MNRAS.422.1122I}; 
		SN 2012A  -- 55933.5, 9.8, 0.04; \citet{2013MNRAS.434.1636T};
		SN 2012aw -- 56002.6, 9.9, 0.07; \citet{2013MNRAS.433.1871B};
		SN 2013ab -- 56340.0, 24.0, 0.04; \citet{2015MNRAS.450.2373B};
		SN 2013by -- 56404.0, 14.8, 0.19; \citet{2015MNRAS.448.2608V};
		SN 2013ej -- 56497.3, 9.6, 0.06; \citet{2015ApJ...806..160B};
		SN 2013hj -- 56637.0, 28.2, 0.10; \citet{2016MNRAS.455.2712B};
		SN 2014G --  56669.7, 24.4, 0.25; \citet{2016MNRAS.455.2712B};
		PTF10iam --  55342.7, 453.35, 0.19; \citet{2016ApJ...819...35A};
		SN 2011fe -- 55797.2, 6.79, 0.023; \citet{2013NewA...20...30M};
		ASASSN-15no -- 57235.5,153.5, 0.045; \citet{2018MNRAS.476..261B}.}
	\label{fig:lc.abs}
\end{figure*}

We further compare \sn\ with a sample of well-studied Type II-P/L SNe in Fig.~\ref{fig:lc.abs}. The slope of the SN is comparable to the slope during the photospheric phase of SNe 2014G (2.55 \maghundred), 2013by (2.01 \maghundred) and 2000dc (2.56 \maghundred) \citep{2016MNRAS.455.2712B}, all of which are fast declining Type II SNe a.k.a SNe II-L. All Type II SNe light curves in the comparison sample show distinct photospheric and radioactive tail phases, with a transition near $ 80-120 $ days. For qualitative comparison, we also include the absolute \textit{R} band light curve of PTF10iam \citep{2016ApJ...819...35A} and the \textit{g} band light curve of ASASSN-15no \citep{2018MNRAS.476..261B} in Fig.~\ref{fig:lc.abs}. PTF10iam had an absolute magnitude similar to \sn\ ($ \sim-20 $ mag) and a somewhat slower decline rate of 2.32 \maghundred. PTF10iam is characterized as a luminous and rapidly rising SN II. \sn\ may have risen equally rapidly, but there is insufficient pre-peak data to be certain. PTF10iam was only observed for $ \sim90 $ days, which is a typical timescale for the photospheric phase to have a nearly constant decline rate, so it is not possible to determine if it had a unique, long-lived linear decline similar to \sn. SN 1979C, SN 1998S and ASASSN-15no had a maximum brightnesses close to, albeit a few tenth mag fainter than, that of \sn, and they also have comparable late-time light curve decline rates to \sn. However, the light curves of SN 1979C, SN 1998S and ASASSN-15no show prominent breaks near $ \sim90 $ days, and so they do not exhibit the most distinguishing feature, the long and continuous linear decline of \sn.

% fig:lc.abs
%%____________________________________________________________________________

The characteristic radioactive tail phase of normal SNe II ($\gtrsim 150 $ days), powered by \cobalt\ to \iron\ decay, has a decline slope of $ \sim0.98~\maghundred$ when the $ \gamma $-ray photons are fully trapped. During this phase, most SNe~II in Fig.~\ref{fig:lc.abs} shows a decline rate consistent with almost full trapping of the $ \gamma $ rays. The light curve of \sn\ is significantly steeper and it continued to decline with a constant slope after the early peak. In comparison, the tail of the light curve of the prototypical Type Ia SN 2011fe \citep{2013NewA...20...30M}, which is also powered by \cobalt\ to \iron\ radioactive decay, has a slope comparable to \sn. The substantially lower $\gamma $-ray optical depth in SNe Ia increases the fraction of escaping $ \gamma $-ray photons and makes the light curve steeper than typical SNe II. The similarity in light curve shapes between SNe Ia and luminous SNe IIL such as SN1979C has been discussed before \citep{1985AJ.....90.2303D, 1987txra.symp..402W, 1989ApJ...342L..79Y}, pointing to the possibility that luminous SN IIL might also be powered by the decay of a large mount of $^{56}$Ni like SNe Ia. We explore this possibility for \sn\, in \S\ref{sec:lc.bol} and  \S\ref{sec:ni}.

\begin{figure}
	\centering
	\includegraphics[width=8.75cm]{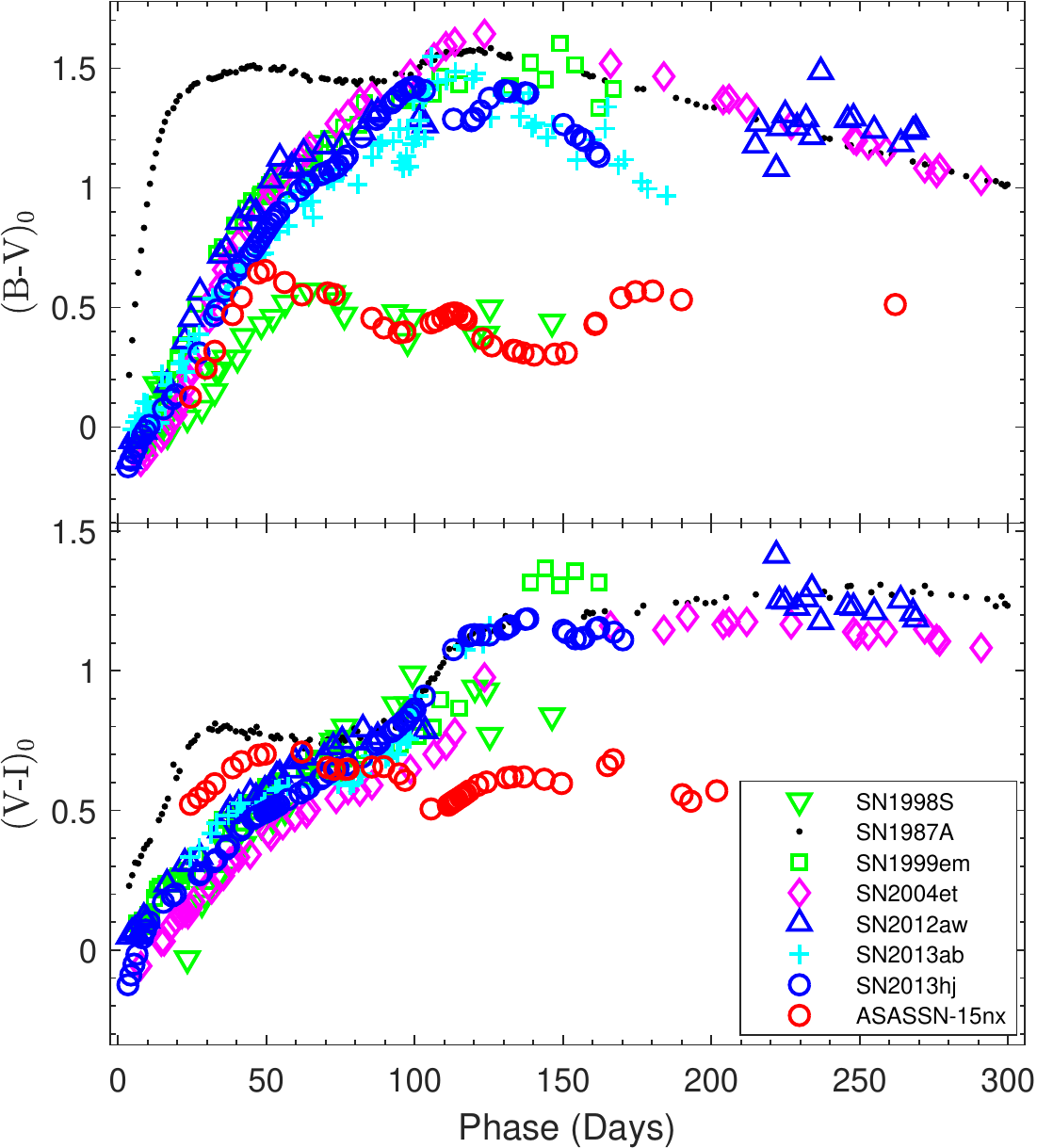}
	\caption{The color evolution of \sn\ as compared to the well-studied  
		Type II SNe 1987A, 1999em, 2004et, 2012aw, 2013ab, 2014hj and also the type IIL/n SN 1998S. The references for the data are same as in Fig.~\ref{fig:lc.abs}.}
	\label{fig:color}
\end{figure}

In Fig.~\ref{fig:color} we present the extinction corrected $ (B-V)_0 $ and $ (V-I)_0 $ color evolution of \sn\, as compared with several SNe IIP/L and the Type IIL/n SN~1998S. 
For \sn, the magnitudes are loosely interpolated to match corresponding photometric epochs for each pair of bands, and this also serves to reduce the random fluctuations from internal uncertainties in the photometry.
For $t <50 $d \sn\ continues to become redder, following a similar trend to other SNe II.  
But after +50d, \sn\ shows a very little change in color, with mean values $ (B-V)_0=0.450\pm0.016 $ and  $ (V-I)_0=0.598\pm0.009 $ mags, in contrast typical SNe IIP/L continues to evolve to substantially redder colors. However, SN IIL/n SN~1998S shows a color evolution similar to \sn\ in terms of the mean colors and the absence of evolution after $ \sim60 $ days.

% fig:color
%____________________________________________________________________________

\begin{figure}
	\centering
	\includegraphics[width=8.75cm]{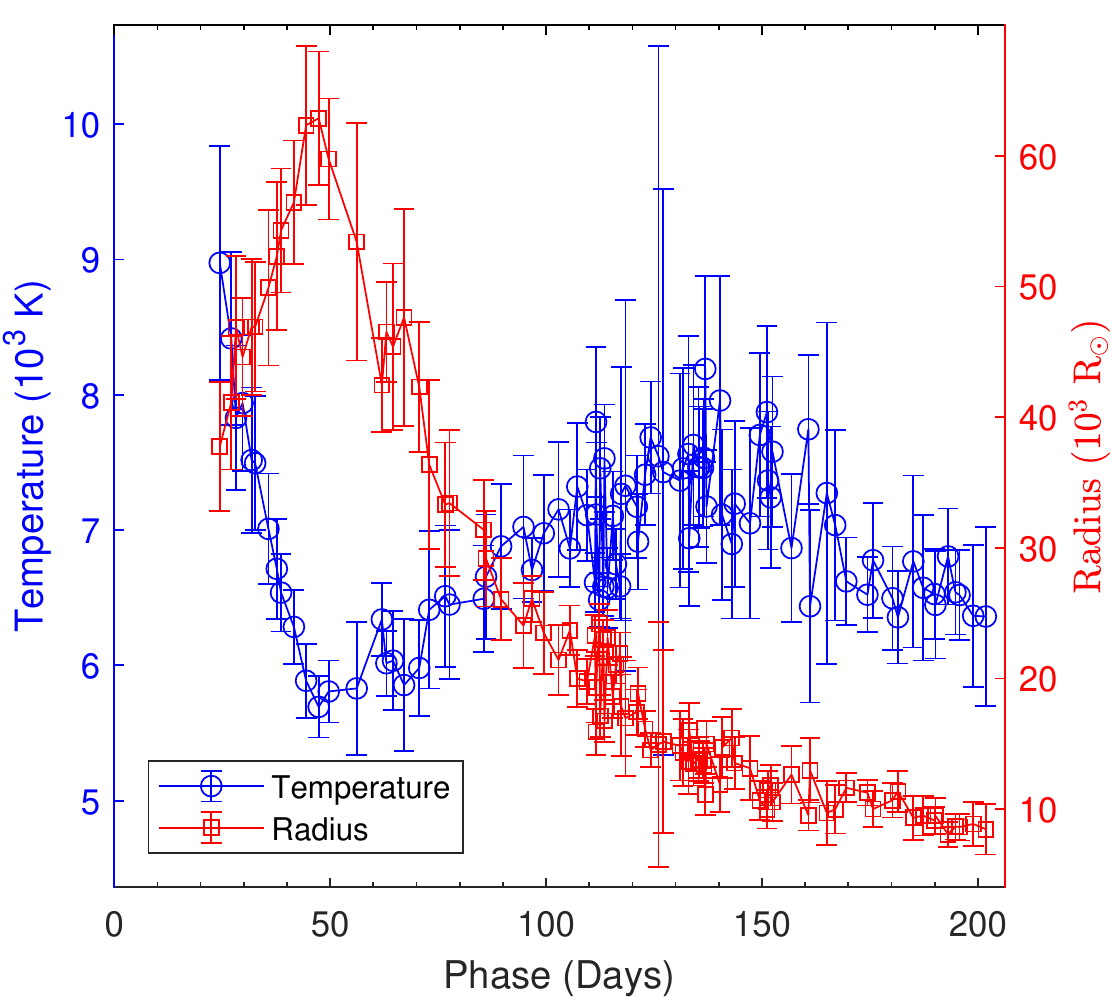}
	\caption{The temporal evolution of the blackbody temperature (blue, left scale) and radius (red, right scale) of \sn.}
	\label{fig:T_R}
\end{figure}

We fit blackbody models to the extinction corrected \textit{B}-, \textit{V}- and \textit{I}-band magnitudes. Fig.~\ref{fig:T_R} shows the resulting evolution of the effective temperature and radius. 
Initially the radius increases until $ +50 $d, where the photospheric phase ends and the ejecta becomes optically thin. During this phase the temperature shows a steady decline as the ejecta cools down with expansion. The best-fit effective temperature then starts to rise until $ \sim135 $d before again beginning a slow decline. This blackbody temperature evolution is unlike that of normal SNe, where we usually simply see monotonic declines \citep[e.g.,][]{2013MNRAS.433.1871B,2015MNRAS.450.2373B}.

% fig:T_R
%____________________________________________________________________________

\begin{figure}
	\includegraphics[width=8.75cm]{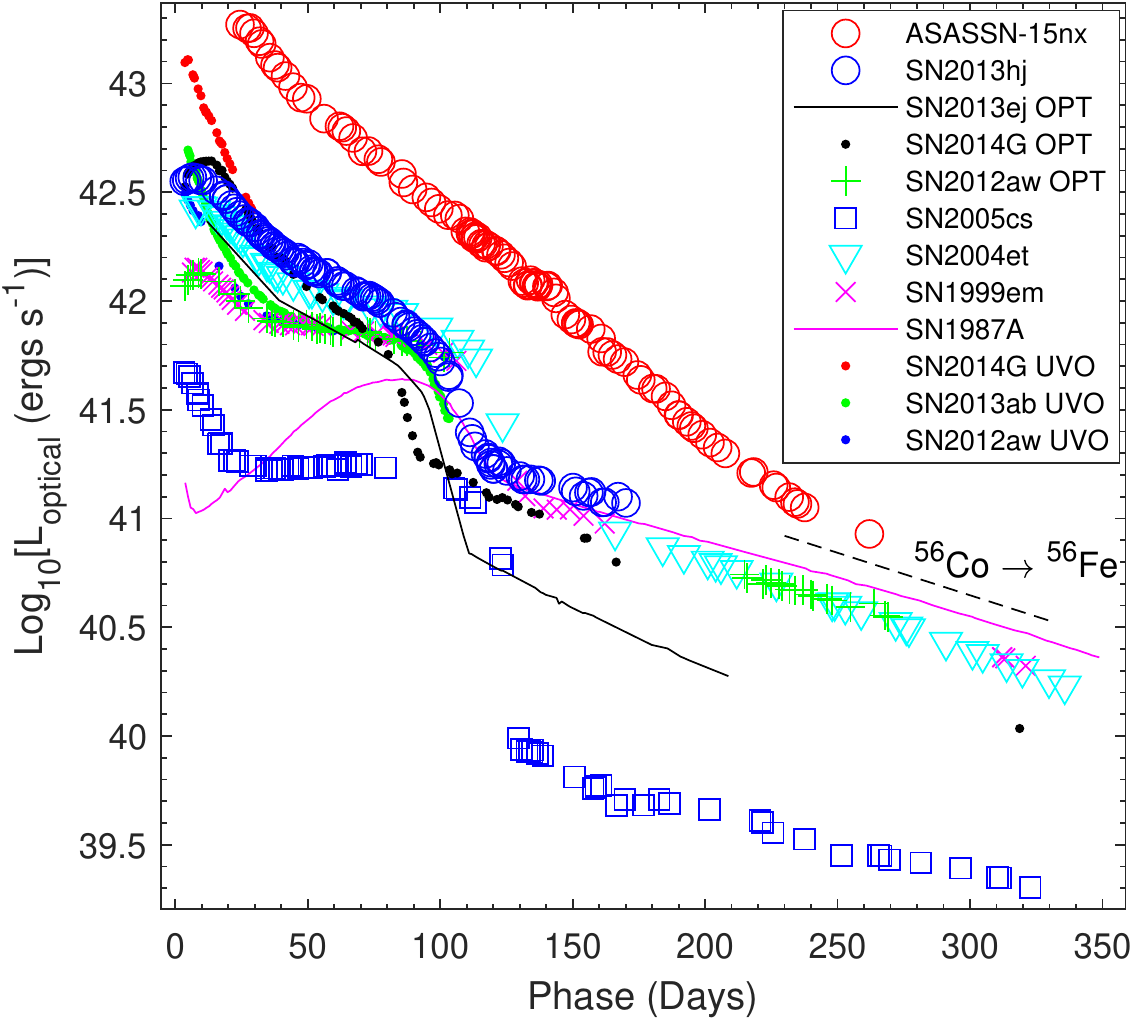}%
	\caption{The \textit{UBVRI} pseudo-bolometric light-curve of {\sn} as compared to other well-studied SNe. Light
		curves including UV contributions are also shown for SNe 2012aw, 2013ab and 2014G (labeled as UVO).
		The adopted distances, reddenings and explosion times are the same as in Fig.~\ref{fig:lc.abs}. The slope of a light curve powered by the radioactive decay of \cobalt$ \rightarrow $\iron\ is shown with the dashed line.}
	\label{fig:lc.bol}
\end{figure}

 %____________________________________________________________________________________________

 \subsection{Bolometric light curve and \rm{\nickel} mass} \label{sec:lc.bol}

In Fig.~\ref{fig:lc.bol} we compare the pseudo-bolometric ($ 3335-8750 $ \AA) light curves of \sn\ with a sample of well-studied SNe II. The pseudo-bolometric luminosities for all SNe in the sample are constructed following the method described in \cite{2013MNRAS.433.1871B}. The light curve decline rate for \sn\ is $ 1.05 $ dex $ (100 \rm~d)^{-1} $, which is $ \sim3 $ times faster than that expected for a light curve powered by the  radioactive decay of \cobalt\ to \iron.

\begin{figure}
	\centering
	\includegraphics[width=\linewidth]{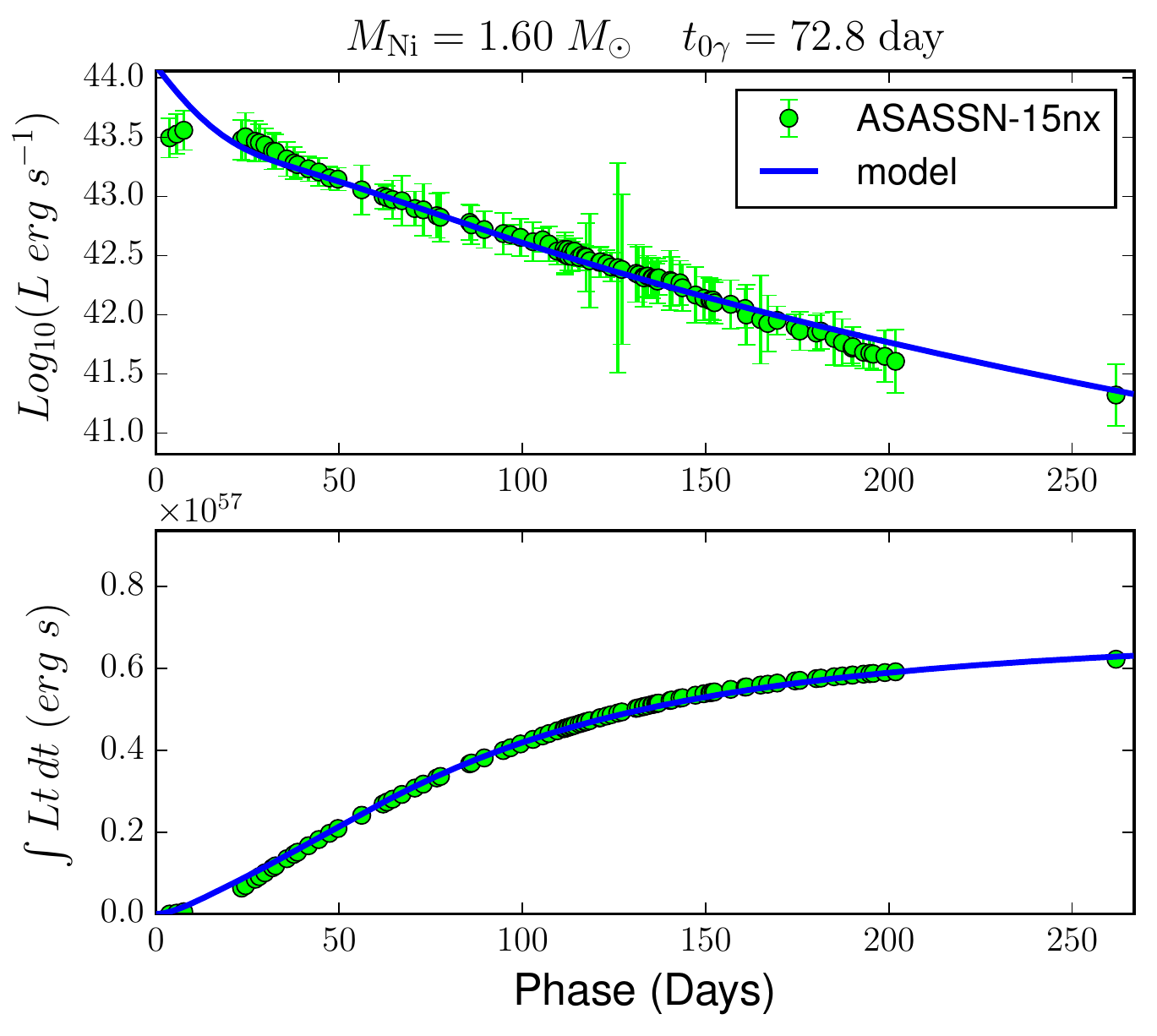}%
	\caption{
	Radioactive decay models of the bolometric light curve (top panel) and the time-weighted integrated radiated energy (bottom panel) of \sn.}
	\label{fig:model}
\end{figure}

Next we modeled the blackbody bolometric luminosity using a pure radioactive $\nickel\rightarrow\cobalt\rightarrow\iron $ decay model. 
The two free model parameters are the \nickel\ mass $ M_{Ni} $ and the $ \gamma $-ray trapping parameter $ t_{0\gamma} $, which defines the evolution of the $ \gamma $-ray optical depth as $ \tau_g\sim t_{0\gamma}^2/t^2 $. While all the positron kinetic energy from \cobalt\ decay is trapped, only $ [1-{\rm exp}(-t_{0\gamma}^2/t^2)] $ fraction of the $ \gamma $-ray decay energy is trapped in the envelope \citep[see the discussions of this approximation in, e.g.,][]{1997ApJ...491..375C,2012ApJ...746..121C}.
As shown in the top panel of Fig.~\ref{fig:model}, 
we find a reasonable fit for $M_{Ni}=1.6\pm0.2 $\msun\ and a $ \gamma $-ray trapping factor $ t_{0\gamma}=73\pm7 $ days. Although the overall light curve is matched well by the model, there are some noticeable deviations at both early ($<25 $ days) and late ($\sim200 $ days) times. Even if the light curve is entirely powered by the radioactive decay, this simple model is not expected to fully capture the light-curve evolution at early phases when the ejecta is optically thick and diffusion is important. The deviation at late time might reflect inaccuracies either in the physical assumptions of the simple radioactive decay model or the blackbody model used in deriving the bolometric light curve.

To gain more insight into the powering source, we also used the time-weighted integrated luminosity method to model the light curve \citep{2013arXiv1301.6766K, ET, bolometricia}. The results are shown in the bottom panel of Fig.~\ref{fig:model}. 
By comparing time-weighted integrated luminosity with a radioactive decay model during the post-photospheric phase, we can verify the powering source of the entire light curve. \cite{2013arXiv1301.6766K} showed that the time weighted integrated luminosity $ (\int_{0}^{t}L(t^\prime)t^\prime dt^\prime) $ can also be used to put an additional constraint on the \nickel\ mass during the post-photospheric phase provided the light curve is solely powered by radioactive decay, as in SNe Ia. 
For this purpose early time estimates of the luminosity are required, so we extrapolated the temperature into the  pre-peak phases and by scaling the SED to the \textit{V}-band flux at fixed temperature. The blackbody luminosity is listed in Table~\ref{tab:BB}.
The uncertainty introduced by the extrapolation will eventually become smaller at larger values of $ t $ due to the time weighting of the integral. 
The comparison of the integrated time weighted luminosity shows that the total energy budget of \sn\ is consistent with almost fully radioactive decay energy, which raises the possibility that it is the dominant powering source of the light curve.

%____________________________________________________________________________
\begin{figure}
	\centering
	\includegraphics[width=0.9\linewidth]{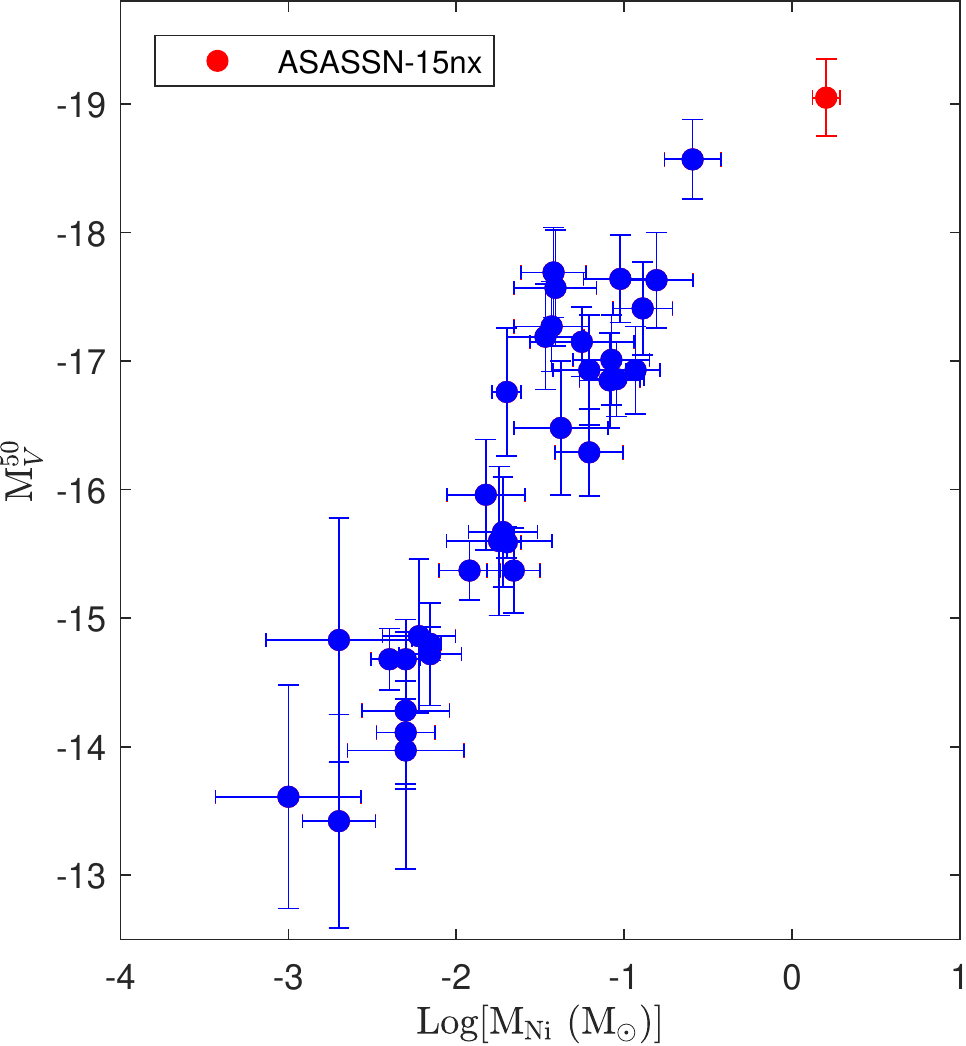}
	\caption{The absolute $ V $ band magnitude at $ t=50 $ days as a function of the estimated \nickel\ mass for the SNe II sample, from \citet{2003ApJ...582..905H} and \citet{2014MNRAS.439.2873S}. \sn, shown with filled red circle, lies on the extrapolation of the correlation to higher masses and luminosities.}
	\label{fig:nicomp}
\end{figure}

In Fig.~\ref{fig:nicomp} we show the correlation between the \nickel\ mass and the \textit{V}-band luminosity $ M_V^{50} $ at $ t=50 $ days for the  34 Type II SNe from \cite{2003ApJ...582..905H} and \cite{2014MNRAS.439.2873S} along with \sn. As one would expect, \nickel\ mass increases with luminosity, however with considerable scatter in the correlation \citep[see][]{2015ApJ...806..225P}. \sn\ is consistent with an extrapolation of this correlation, while it clearly stands out at the higher extreme end of the luminosity distributions.

The black body model may not always be a good approximation, especially in the optical thin phases. 
We also constructed the bolometric light curve using bolometric corrections derived empirically from other well observed Type II SNe.  The correction factor is calibrated as a function of broadband color. We adopt the bolometric correction from \cite{2009ApJ...701..200B} based on the $ (B-V) $ color. Since \textit{B}-band data is not available during the pre-peak phases, the $ (B-V) $ color is linearly extrapolated from the color curve for $ t<50 $ days. By modeling this bolometric light curve we estimate $M_{Ni}=1.3\pm0.2 $\msun\ and $ t_{0\gamma}=71\pm10 $ days. This value of nickel mass is consistent within $ \sim20\% $ to that estimated with our earlier model, and the $ \gamma $-ray trapping factor is also similar. We caution that the bolometric luminosities used in this method may also be inaccurate for an SN like \sn\ whose light curve, color and spectroscopic evolution is significantly different from generic SNe II.

% sec:sp
%________________________________________________________________________________

\section{Optical spectra} \label{sec:sp}

% fig:sp.all
%____________________________________________________________________________
\begin{figure*}
	\centering
	\includegraphics[width=\linewidth]{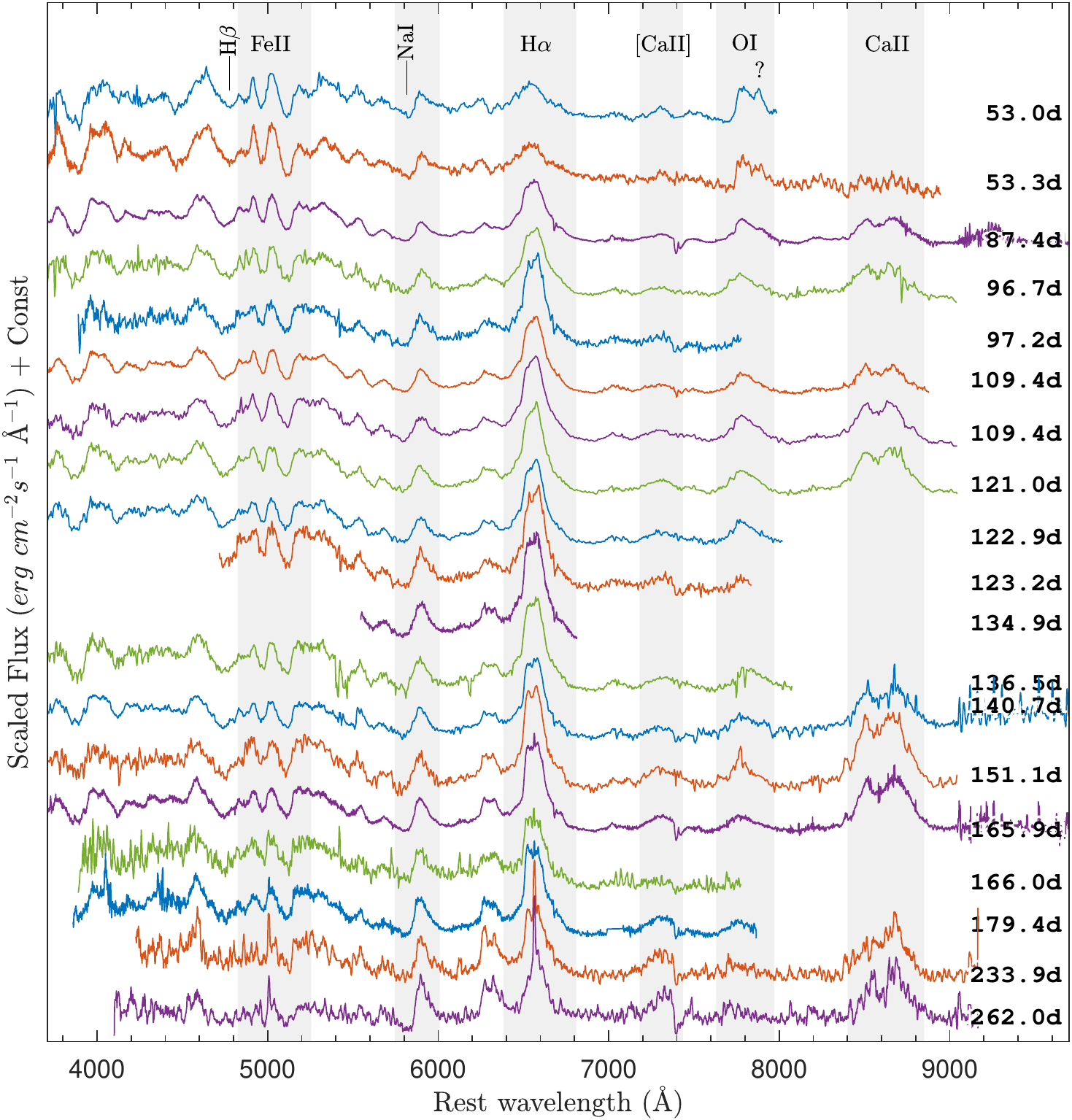}
	\caption{The rest-frame spectral evolution of \sn\ ordered by the age with respect to the explosion epoch \EpEpoch. Prominent lines of hydrogen (\ha, \hb), iron (\Feii\ \ldld~4924, 5018, 5169), sodium (\Nai\ \ld5890), calcium and oxygen are marked. Spectra with low SNR have been binned in wavelength to reduce the noise.}
	\label{fig:sp.all}
\end{figure*}

\subsection{Key spectral features} \label{sec:sp.key}

The spectroscopic evolution of \sn\ fom $ t=53 $ to 262d is presented in Fig.~\ref{fig:sp.all}.
A striking feature is the \ha\ emission profile, which has an unusual triangular shape in the earliest spectra at $ t=53$d. We discuss the \ha\ profile and its evolution further in \S\ref{sec:sp.comp} and \S\ref{sec:sp.evolution}.
Forbidden [\Caii] (\ldld 7291, 7324) emission, which usually becomes visible not until the early nebular phase ($ \gtrsim 120 $ days), is visible throughout our observing campaign. The \Caii\ multiplets (\ldld 8498, 8542 and 8662) are not detected at $t=53.3$d, though the SNR in the relevant part of the spectrum is relatively low, while they are clearly visible at $t=87.4$d, which is typical of SNe IIP/L. \Oi\ emission near 7780\AA\ is another prominent feature in the 53d spectra and is present throughout the spectral evolution. The strong \Oi\ emission feature appears unusually early for a Type II SN. Interestingly, the \Oi\ line has a doubly-peaked profile in the 53d spectra, which evolves into a singly-peaked profile by 96.7d and is no longer present in all subsequent spectra. The origin of the redder component of the double-peaked profile is not clear (as marked with a question mark in Fig.~\ref{fig:sp.all}).
The overall spectral appearance and the early presence of [\Caii] and strong \Oi\ features in 53.0d make the spectrum appear much more evolved than typical SNe IIP/L at a similar phase. 
Another noticeable feature in all the spectra is the apparent continuum break near 5500\AA. The continuum level on the bluer side is higher by about 25\% than the red side (see further discussions in the context of \synow\ modeling below and also in \S\ref{sec:csm}). The evolution of the prominent metallic lines of \Feii\ (\ldld~4924, 5018, 5169) and \Nai~D (5893 \AA) is typical of SNe II.
 
 % fig:sp.synph 
 %____________________________________________________________________________
 \begin{figure*}
 	\centering
 	\includegraphics[width=0.9\linewidth]{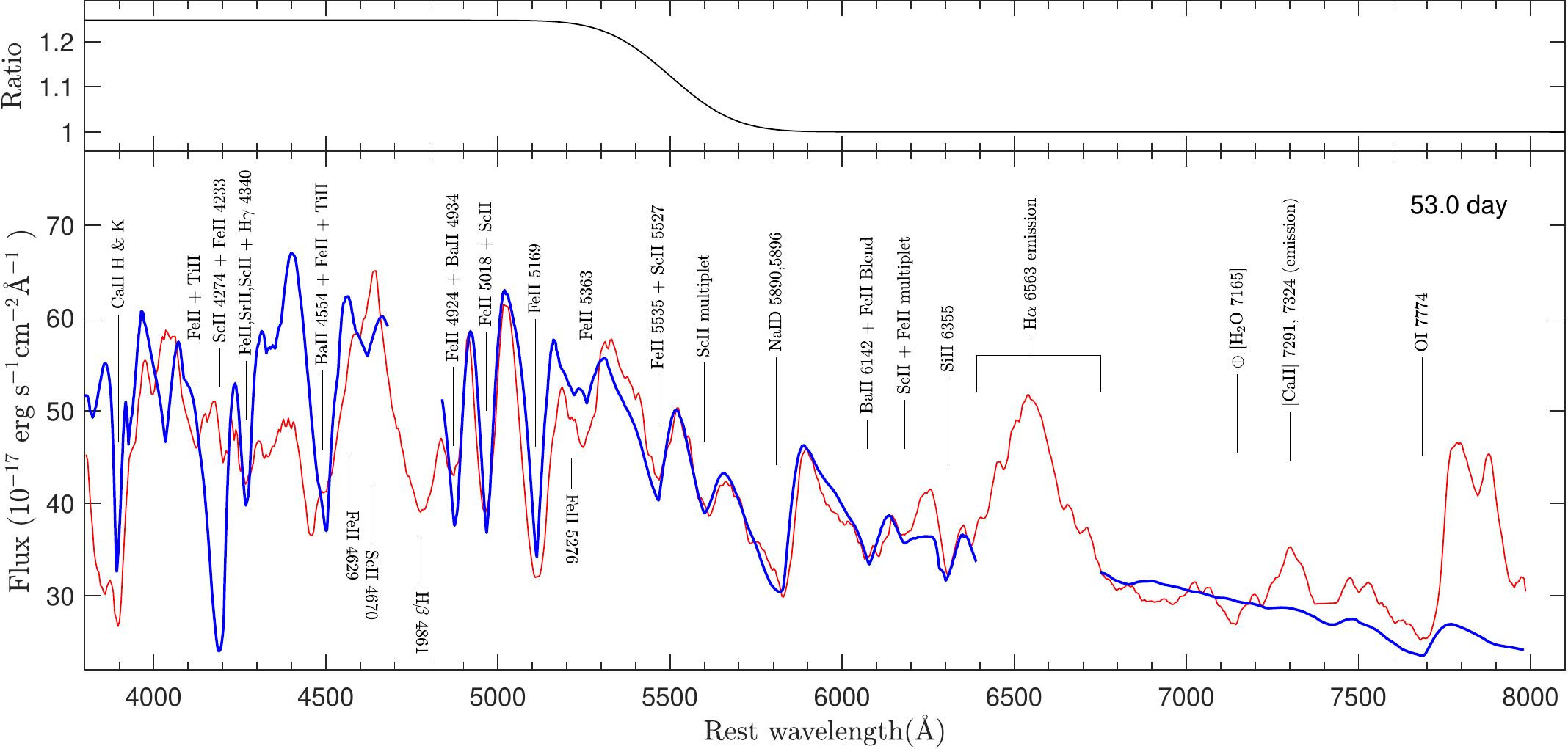}%
 	\caption{\synow\ model (blue) of the $ t=53.0 $~day spectrum of \sn\ (red). Observed fluxes are corrected for extinction. All \sn\ spectra show a break in the continuum near 5500\AA, where the blue side has a higher flux level. To reproduce this feature, the model continuum is multiplied by the Gaussian convolved step function shown in the top panel.}
 	\label{fig:sp.synph}
 \end{figure*}

% sec:sp.syn
%________________________________________________________________________________-

We used \synow\footnote{https://c3.lbl.gov/es/\#id22} \citep[e.g. ][]{1997ApJ...481L..89F,1999MNRAS.304...67F,2002ApJ...566.1005B} to model the spectra and identify features.
In order to mimic the continuum break, 
we multiply the model spectrum with a Gaussian convolved step function, whose amplitude and width are tuned to fit the observed spectrum. This modifier function is shown in the top panel of Fig.~\ref{fig:sp.synph}, where the bluer continuum beyond 5500\AA\ is $ \sim25\% $ higher than the red with a 150\AA\ width for the convolving Gaussian function. 

The set of atomic species used to generate the synthetic spectrum are 
\Hi, \Hei, \Oi, \Feii, \Tiii, \Scii, \Caii, \Baii, \Nai\ and \Siii. As  noted above, \sn\ exhibits an unusual, triangularly-shaped \ha\ emission with a weak absorption feature on the blue side is not consistent with a P-Cygni profile. The \hb\ profile is also unusually broad and extended, which \synow\ could not reproduce with a single \hb\ component using any combination of expansion velocity and optical depth profile. So, the \ha\ and \hb\ line region have been masked in the synthetic spectrum. Apart from the nebular-like emission features of \Caii\ and \Oi, for which SYNOW is not applicable, many of the spectral features are reasonably well reproduced except for the \Scii\ features at 4274\AA\ and 4670\AA.

In the \synow\ model, the absorption feature near 6300\AA\ which appears to form the absorption component of the P-Cygni profile associated with \ha, is reproduced by \Siii\ (6355\AA). The \Siii\ velocity is same as the photospheric velocity ($ \sim3.3\kkms $)
found for the other metal lines, affirming the line identification and suggesting that there is little or no absorption component associated with \ha.

% fig:sp.lit 
%____________________________________________________________________________
\begin{figure}
	\centering
	\includegraphics[width=\linewidth]{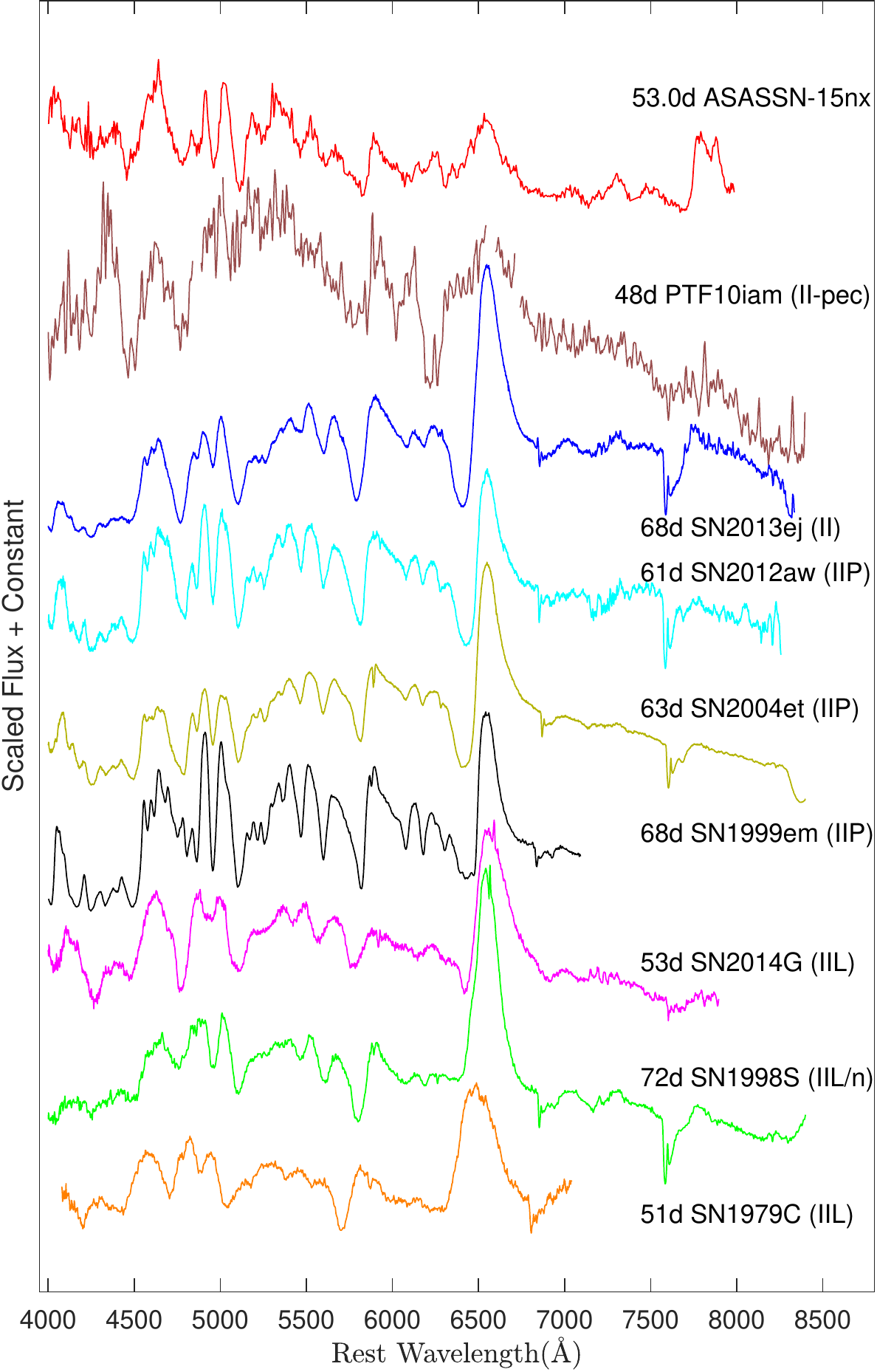}%
	\caption{Comparison of the 53.0d spectrum of {\sn} to other Type II SNe 2013ej \citep{2015ApJ...806..160B}, 2012aw \citep{2013MNRAS.433.1871B}, 2004et \citep{2006MNRAS.372.1315S}, 1999em \citep{2001ApJ...558..615H}, 2014G \cite{2016MNRAS.462..137T}, 1998S \citep{2000ApJ...536..239L,2001MNRAS.325..907F}, PTF10iam \citep{2016ApJ...819...35A} and 1979C \citep{1981ApJ...244..780B} at similar age. All 
		comparison spectra are corrected for extinction and redshift. For PTF10iam, the regions contaminated by host emission lines are masked.}
	\label{fig:sp.comp54}
\end{figure} 

% fig:sp.lit 
%____________________________________________________________________________
\begin{figure}
	\centering
	\includegraphics[width=\linewidth]{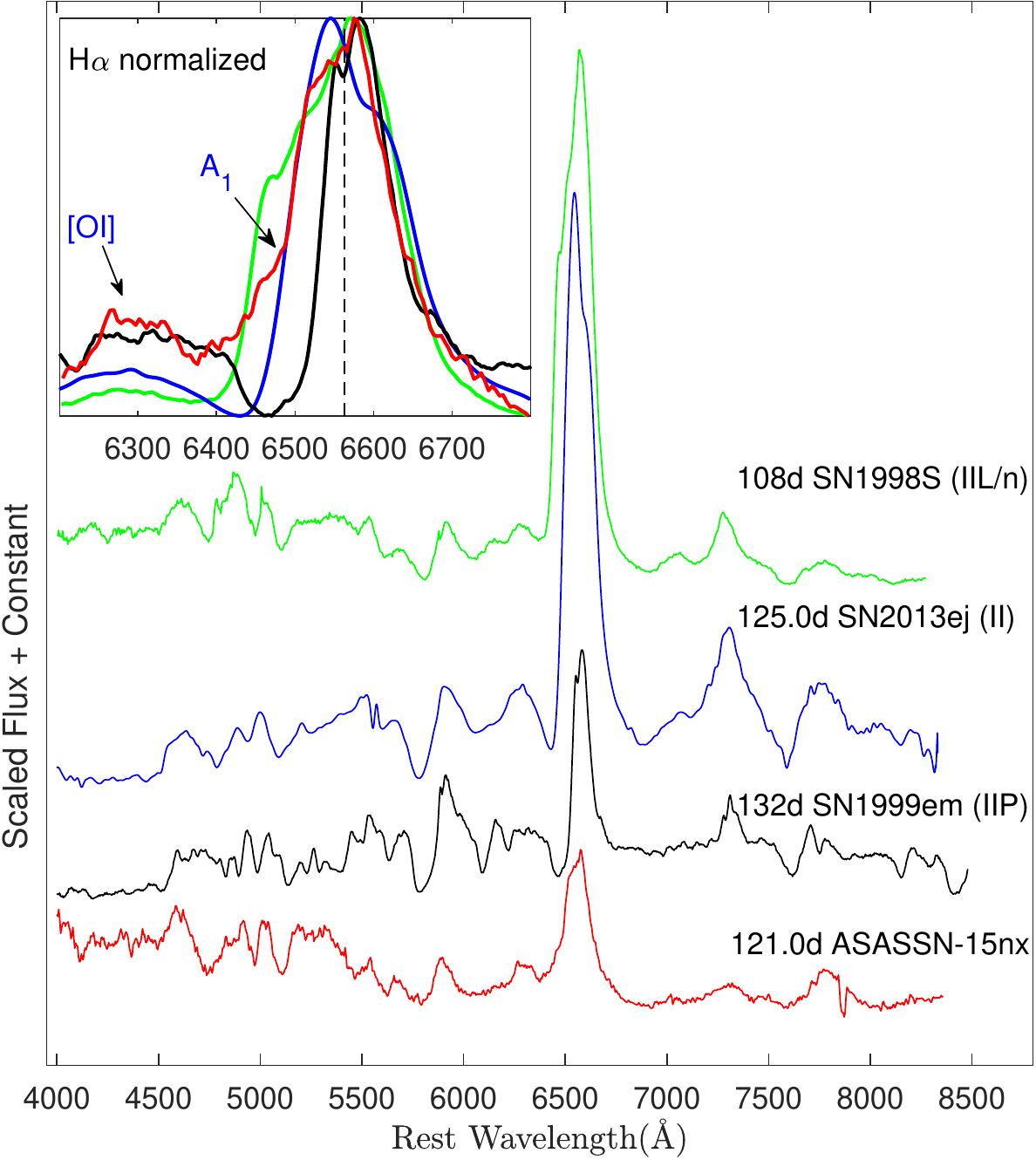}%
	\caption{Same as Fig.~\ref{fig:sp.comp54}, but for the 121.0d \sn\ spectrum. The inset shows the \ha\ region only, scaled to the peaks of the \ha\ profiles.}
	\label{fig:sp.comp124}
\end{figure} 

% fig:sp.neb 
%____________________________________________________________________________
\begin{figure}
	\centering
	\includegraphics[width=8.5cm]{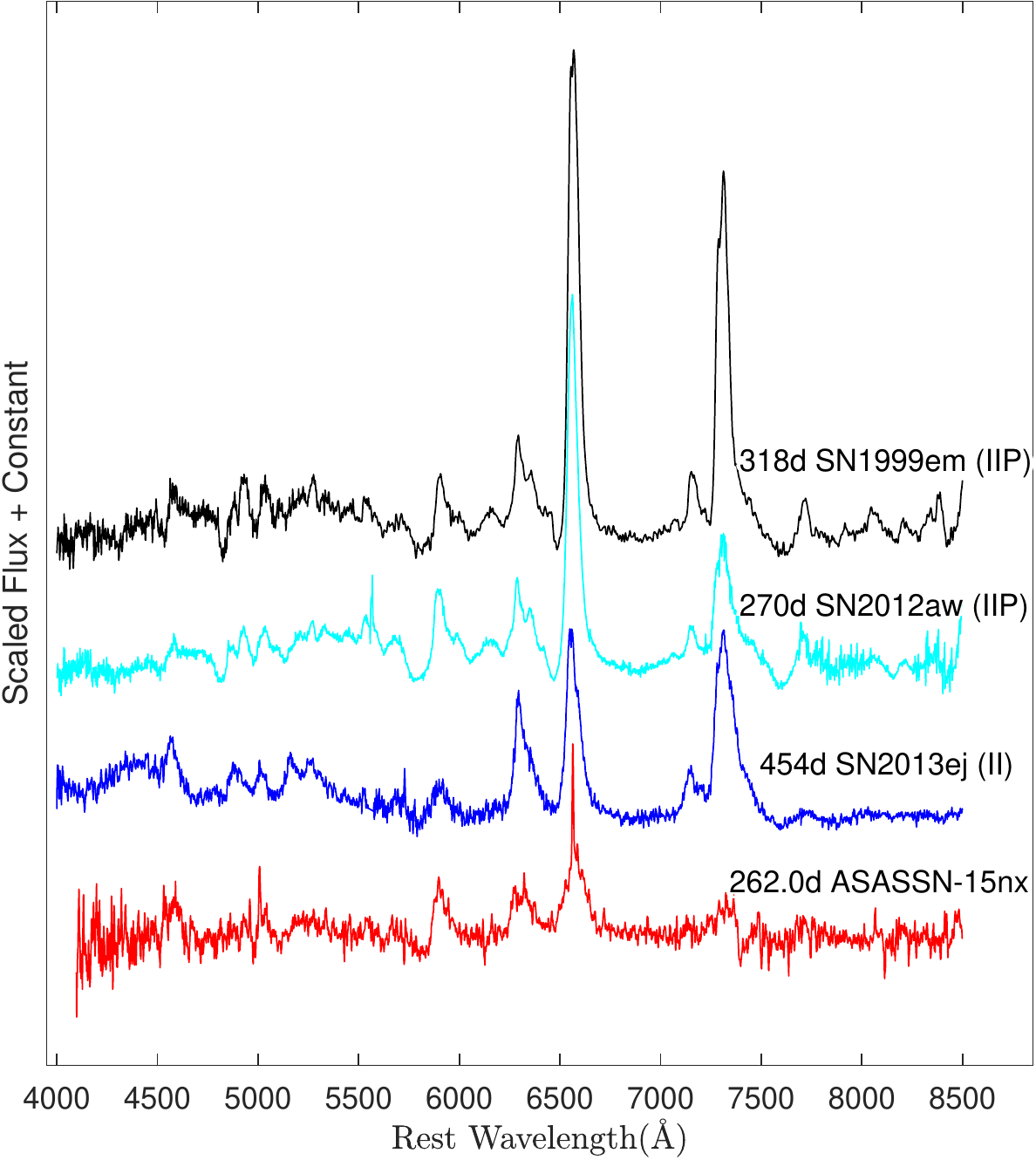}%
	\caption{Same as Fig.~\ref{fig:sp.comp54}, but for a late phase (262.0d) spectrum of {\sn}.}
	\label{fig:sp.comp269}
\end{figure}

\subsection{Comparison of Spectra} \label{sec:sp.comp}
We compare spectra of \sn\ to other Type II SNe at three different phases (53.0d, 121.0d, 262.0d) as shown in Figs.~\ref{fig:sp.comp54}--\ref{fig:sp.comp269}. 
Fig.~\ref{fig:sp.comp54} shows the comparisons of the 53.0d spectrum. Apart from the \ha\ profile, the spectral features broadly resemble most of the other SNe~II spectra in the comparison sample. The \ha\ profile of \sn\ is significantly weaker and unusually triangular in shape as compared to other SNe II. For instance, normal SNe~II at $ t\approx53 $ days have a mean \ha\ equivalent width of $ \sim157 $\AA\ \citep{2017ApJ...850...89G}, whereas for \sn\ we find a significantly lower value of $ \sim117 $\AA\ and it stays systematically weak throughout the evolution. P-Cygni \ha\ profiles with an absorption component are common for most SNe II at this phase. But the absorption component of the \ha\ profile appears to be non-existent for \sn, as in the \synow\ model discussed above where the absorption feature blue-ward to the \ha\ emission is identified as \Siii. The weakening of the \ha\ absorption component has been generally seen for luminous and fast declining SNe IIL \citep[see e.g.,][]{2014ApJ...786L..15G} such as SN2014G, SN1998S and SN1979C shown in Fig.~\ref{fig:sp.comp54}, and the absence of this absorption component in \sn\ is consistent with this trend. 
The \ha\ profile for \sn\ also has an unusual triangular shaped peak
compared to other known SNe II including SNe II-L. 
The doubly-peaked and exceptionally strong \Oi\ feature near 7700\AA\, is also not seen in any of the comparison spectra. 
The spectrum of PTF10iam, whose early-phase light curve closely resembles \sn\ (see Fig.~\ref{fig:lc.abs}) is also shown in Fig.~\ref{fig:sp.comp54}. Apart from showing a weak and irregularly shaped \ha\ profile, the spectrum of PTF10iam does not show any other similarities with \sn.

In Fig.~\ref{fig:sp.comp124} we compare the 121.0d spectrum with three other SNe II spectra at a similar age.  The comparison spectra are specifically selected to show some form of irregularities or unusual shapes in their \ha\ profiles. The \sn\ spectrum still has weaker \ha\ emission. In the inset of the Fig.~\ref{fig:sp.comp124}, we zoom into the \ha\ line, scaling each spectra to the line peak. Unlike other SNe, the continuum of \sn\ on the blue side of \ha\ has a higher flux level than on the red side. 
SNe 1999em and 2013ej both show asymmetric, possibly doubly-peaked components for \ha. Such profiles are often attributed to a bipolar distribution of \nickel\ in a spherically symmetric hydrogen envelope \citep{2003MNRAS.338..939E,2015ApJ...806..160B}. Similar asymmetric \ha\ emission due to aspherical \nickel\ distribution has also been observed in SNe 1987A \citep{1995A&A...295..129U} and 2004dj \citep{2006AstL...32..739C}. The \ha\ line profile of \sn\ appears to be even more complicated,   and may also be composed of more than one component though not clearly separable. The wedge-shaped peak of \sn\ is similar to that of the Type II-L/n SN 1998S \citep{2000ApJ...536..239L}, although the latter has a broader emission profile. 

In Fig.~\ref{fig:sp.comp269} we compare the 262d spectrum with three other nebular-phase spectra of Type II SNe. \sn\ continues to have a weak and triangular \ha\ profile. The nebular phase features, like forbidden [\Oi] near 6330\AA\ and [\Caii] near 7300\AA, are significantly weaker than in the other SNe. The \Nai~D feature near 5900\AA\ and the marginally visible \Feii\ feature near 5000\AA\ are comparable in strength with those of other SNe II. 

 % fig:sp.line 
 %____________________________________________________________________________
 \begin{figure}
 	\centering
 	\includegraphics[height=21.3cm]{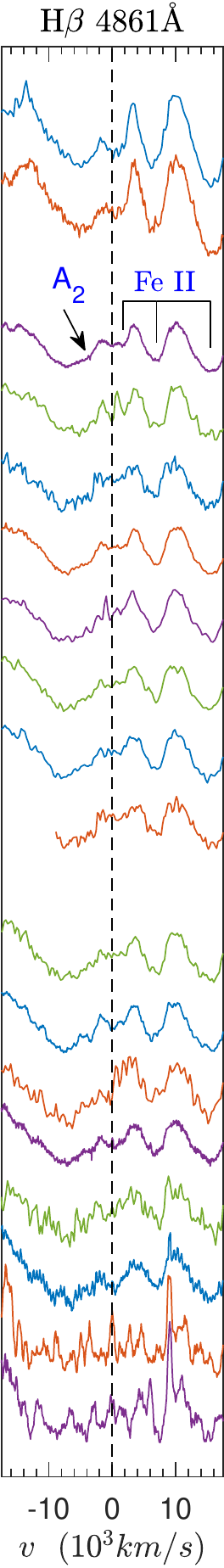}
 	\includegraphics[height=21.3cm]{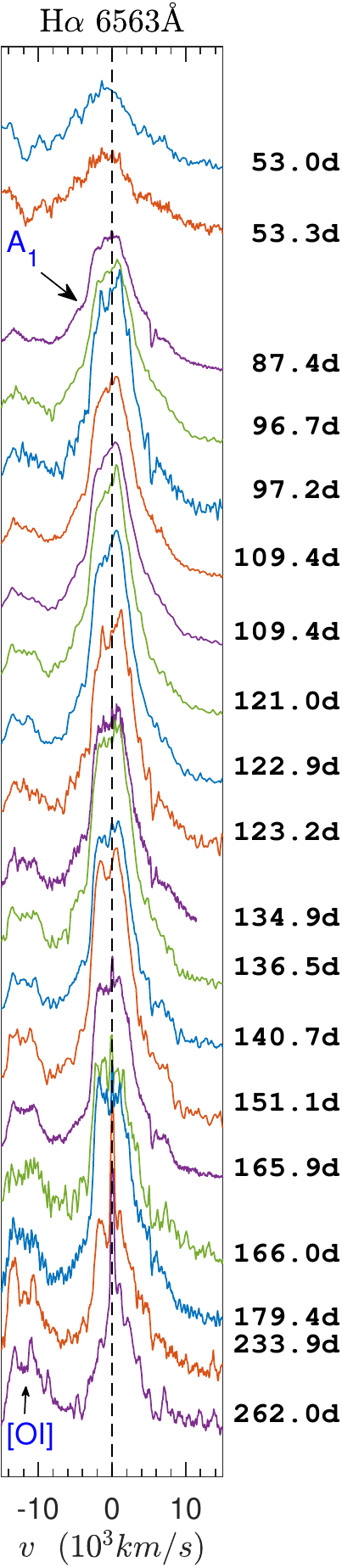} \hskip -4mm%
 	\caption{Spectroscopic evolution in velocity domain corresponding to the \ha\ and \hb\ rest wavelengths. The \Feii\ multiplets are also in the \hb\ window. 
 	}
 	\label{fig:sp.line}
 \end{figure}

\subsection{Evolution of spectral features} \label{sec:sp.evolution}

In Fig.~\ref{fig:sp.line} we show the spectral regions centered on \ha\ and \hb\ in velocity domain. The \Feii\ multiplets (\ldld~4924, 5018, 5169) do not show any significant evolution until the 233.9d spectrum. The \Siii\ (6355\AA) absorption feature identified by \synow\ in the 53d spectra is not detectable from 87d onwards. In the same wavelength range, the [\Oi] emission lines (\ldld~6300, 6364; indicated in the figure) start to appear and become stronger at later times. For a typical Type II SN, the forbidden [\Oi] emission is seen only in nebular phase spectra at $ t\sim150 $d. The \ha\ profiles show a break or an abrupt change in slope on the blue wing of the line profile in all the spectra following 87d. This feature is marked as `A$ _1 $' in Fig.~\ref{fig:sp.line} and in the inset of Fig.~\ref{fig:sp.comp124}. The position of the A$_1  $ feature is  blue-shifted by $ \sim3.60\pm0.25 $\kkms\ from the \ha\ rest frame and remains almost unchanged after it appears at 87d. We also find a similar kink marked as 'A$ _2 $'  in the blue wing of the \hb\ profile. Interestingly, this feature also appeared from 87d onwards and is also blue-shifted by $ \sim3.60 $\kkms\ with respect to \hb. The simultaneous appearance of A$ _1 $ and A$ _2 $  at consistent velocities suggests that the features must have common association with \Hi, rather than any possible blending with other spectral lines. However, the structural configuration of the \Hi\ materials needed to produce such an unusual feature is unclear.

The \ha\ profile shows an atypical triangular shaped emission in the 53d spectra, and then developed a peculiar wedge shaped top at $\sim 87-123 $ days. After this phase, the emission-top shows irregular and possibly multi-component emission features.
The apparent absorption feature near 6400\AA, which can be seen throughout the evolution maybe due to the \Siii\ absorption in the 53d spectra and the [\Oi] emission feature after 87d. This would imply that the \ha\ line lacks the P-Cygni absorption component expected from a typical SN atmosphere. The \hb\ absorption is unusually broad and extended as compared to  other SNe II (see Figs.~\ref{fig:sp.comp54} and \ref{fig:sp.comp124}), which \synow\ model cannot fit with a single \Hi\ component.
It is possible that multiple \hb\ absorption components are blended together to produce the broad feature. 
Such a scenario, with two \Hi\ components resulting in broader \ha\ and \hb\ absorption profiles has been seen in SNe 2012aw \citep{2013MNRAS.433.1871B} and 2013ej \citep{2015ApJ...806..160B}. However, this explanation may not hold for \sn\ due to the missing \ha\ absorption feature.

 % fig:sp.velall
 %____________________________________________________________________________
 \begin{figure}
 	\includegraphics[width=8.5cm]{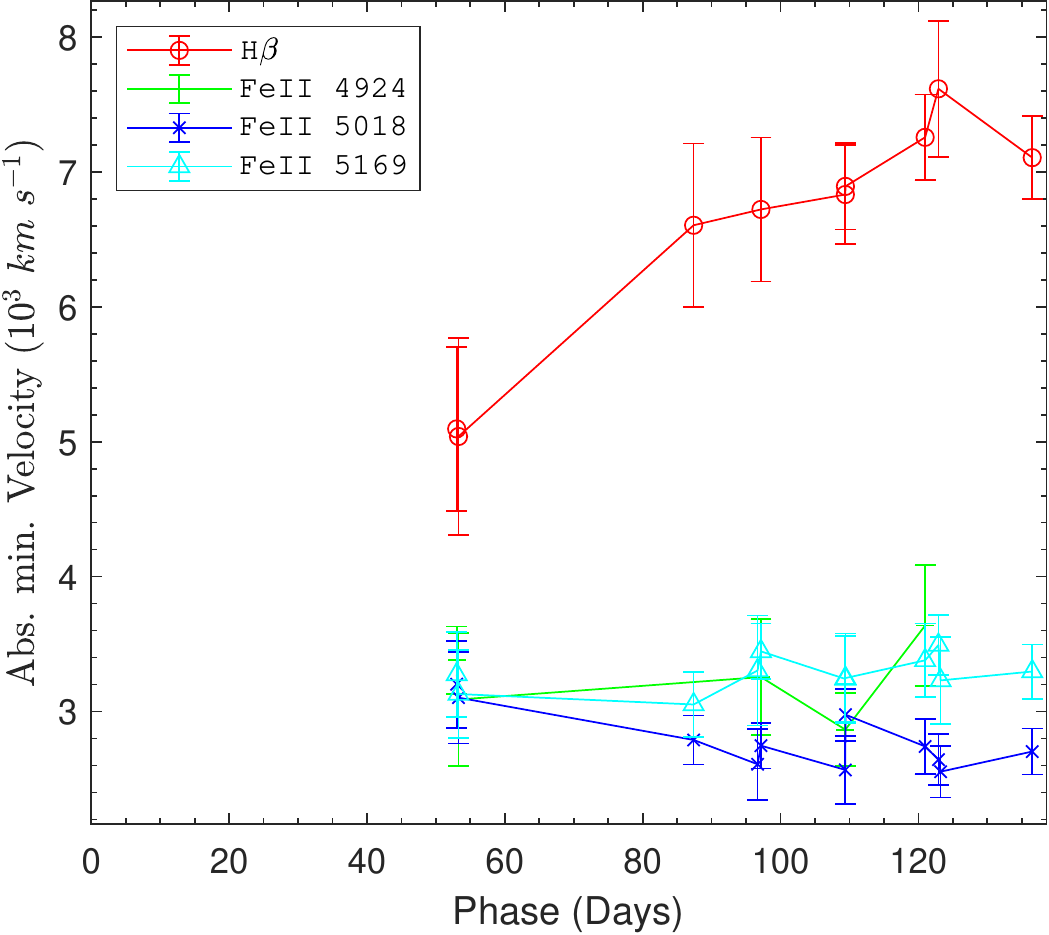}%
 	\caption{Velocity evolution of the \hb, and \Feii\ lines for \sn. The velocities are estimated from the 
 		blueshift of the apparent absorption minima. No attempt has been made to decouple any possible contamination from other lines or high-velocity features. 
	}
 	\label{fig:sp.velall}
 \end{figure}

 % fig:sp.velph
 %____________________________________________________________________________
 \begin{figure}
 	\centering
	\includegraphics[width=8.5cm]{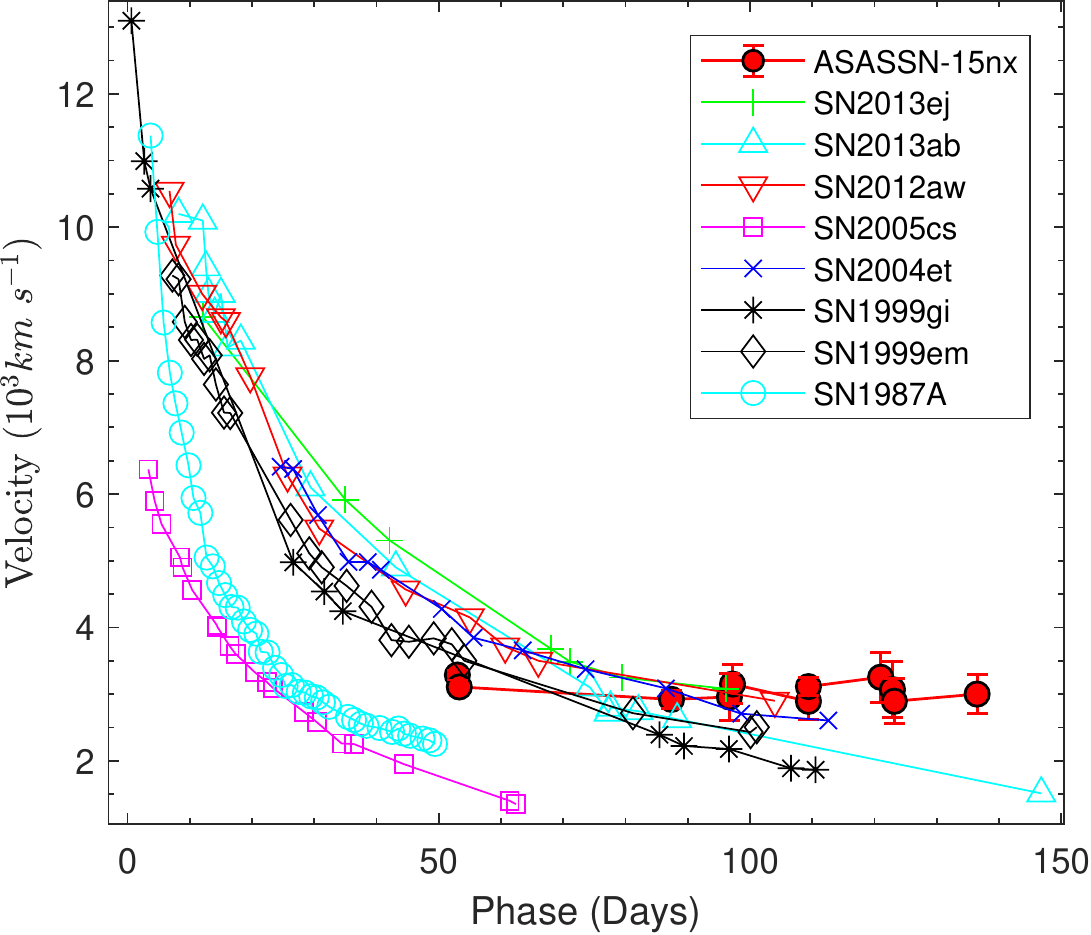}%
 	\caption{The photospheric velocity evolution ($v_{\rm ph}$) of {\sn} as compared with other well-studied Type II SNe.
 		The $v_{\rm ph}$ is estimated by the \Feii\ absorption trough velocities.}
 	\label{fig:sp.velph}
 \end{figure}

Figs~\ref{fig:sp.velall} and \ref{fig:sp.velph} illustrate the spectroscopic velocity evolution of \sn. In Fig.~\ref{fig:sp.velall} we show the \hb\ and \Feii\ velocities using the blue-shifted absorption feature for each line. This includes no corrections for possible blended components. 
The \hb\ velocity evolution is puzzling, rising from  5.0\kkms\ at 53d to $ \sim6.6 $\kkms\ at 87d and then continuing to increase.
The peculiar \hb\ velocity evolution may indicate the presence of high-velocity components within the absorption profiles. An interaction of outer ejecta with CSM can produce such high-velocity components, which are expected to remain almost constant in velocity throughout the evolution \citep{2007ApJ...662.1136C,2015ApJ...806..160B,2017ApJ...850...89G}. If the high-velocity component is strong enough, and the regular \hb\ component continues to become weaker, then the effective minima of the blended trough can show an increasing blueward shift, as we see in \sn. 
On the other hand, the \Feii\ lines, which are generally regarded as good tracers of the photosphere, show almost no variation during the entire spectral evolution.
In Fig.~\ref{fig:sp.velph}, the \Feii\ line velocities of \sn\ are compared with other Type II SNe. Unlike the other SNe, the \Feii\ velocity of \sn\ remains almost  unchanged at $ \sim3.1 $\kkms.

\section {Summary of peculiarities} \label{sec:sum}

\sn\ exhibits a number of unusual features which can be summarized as follows.
\begin{enumerate}
	\item The peak magnitude of $ \sim -20 $ lies in the luminosity ``gap" between normal CCSNe and SLSNe.
	\item It has a uniquely, long-lived, post-peak linear light curve decline with a steep slope of 2.5 \maghundred\ (i.e., an exponential decline in flux $\propto e^{-t/43 {\rm d}}$) until the end. The perfectly linear light curve extended until end of observations at 262d. This is much longer than any known SNe IIP/L, which always show a change in light curve slope at around 100 days after maximum light, marking the transition to a $^{56}$Co radioactive decay tail with a slope of 0.98 \maghundred.
	\item The broadband colors are almost constant and remain blue after 50 days. Equivalently, the blackbody temperature monotonically decreases until 50d as expected, reaching at $T\sim5.8 $kK. However, after this epoch we see an upward trend in the temperature reaching a value of $ \sim7.5 $kK near 135d, before it again starts to decrease. Such an evolution has not been observed in any other SNe~II.
	\item The spectra of \sn\ shows unique, triangular \ha\ emission profile throughout its evolution. The strength of the \ha\ emission is weak  compared to a typical SN II-P/L.
	\item The \ha\ profile shows no evidence for an associated P-Cygni absorption trough. The apparent absorption minima  on the blue wing appears to be due to the presence of a \Siii\ feature at 53d, and then an [\Oi] feature at later times.
	\item Nebular spectral features like \Oi\ (7774 \AA), [\Caii]  (\ldld 7291, 7324) and [\Oi] (\ldld 6300, 6364) appeared earlier than in a typical SN II-P/L. For example, the [\Oi] features in typical SNe IIP appears after $ \sim150 $ days, while in \sn\ they  are detectable starting at 87d.
    Due to the presence of these features, the spectra of \sn\ appear more evolved than a typical SN II at similar age. 	
    \item The \Oi\ (7774 \AA) feature show a double peak emission at 53d. The unidentified redder component is much weaker at 87.4d and completely disappeared in later epoch.
    \item The spectra all show an abrupt continuum break near 5500\AA. The blue continuum is systematically brighter than the red (e.g., by $ \sim 24$\% at 53.0d).
	\item There is a break in the blue wing of the both \ha\ and \hb\ profiles (marked as `A$ _1 $' and `A$ _2 $' in Figs.~\ref{fig:sp.line}~and~\ref{fig:sp.comp124}), starting at 87d  with a blueshifted velocity of $ \sim3.6 $\kkms, indicating that both of these features have a common \Hi\ origin.
	\item The \hb\ velocity shows an unusual increasing trend with time. The \hb\ absorption profile is also unusually broad and extended. This could not be reproduced by a single \Hi\ component in our \synow\ models.
	\item The \Feii\ line velocities show no evolution, while in typical SNe, all line velocities are expected to decay with time.
\end{enumerate}

\section{Discussion} \label{sec:discuss}

We are not aware of a theoretical model that can explain all the peculiarities of \sn. One important question is what powers this luminous SN II and how the long-lasting, ``perfectly'' linear light curve is produced, which are discussed in \S\ref{sec:ni} and \S\ref{sec:csm}. In \S\ref{sec:rate}, we comment on the rate of \sn-like SNe.

\subsection{High \nickel\ mass}\label{sec:ni}

One possible scenario for powering \sn\ is the synthesis of a large amount of radioactive \nickel. As discussed in \S\ref{sec:lc.bol}, matching the high luminosity requires an exceptionally large amount of \nickel, $ M_{Ni}=1.6\pm0.2$ \msun. This is significantly higher than normal Type II SNe which have average $ M_{Ni}$ of 0.05 \msun\ and extend to at most $ \sim0.1-0.2 $ \msun\ \citep[see e.g.][]{2017ApJ...841..127M}. {For stripped envelope SNe, it can be as high as $ 0.6-0.7 $ \msun\ \citep[e.g. for type Ic SN 2011bm;][]{2012ApJ...749L..28V}}.
The best-fit model also requires inefficient $ \gamma $-ray trapping, with $ t_{0\gamma}\approx71 $ days. For comparison, normal SNe IIP are consistent with complete trapping (i.e., $ t_{0\gamma}\to \infty$), and for SNe Ia, $ t_{0\gamma}\approx40 $ days \citep{bolometricia}. The value of $ t_{0\gamma} $ for \sn\ implies that the envelope is inefficient at thermalizing the $ \gamma $-rays. This is also consistent with the weak \ha\ emission, because the hydrogen content in an SN ejecta is the dominant source of the $\gamma $-ray opacity. 
 
 The implied high $^{56}$Ni mass and the short gamma-ray escape time of $t_0\sim 70~\rm day$ constrain the ejecta structure. 
The $ \gamma $-rays must escape the region in the ejecta where the \nickel\ is concentrated allowing a constraint on the iron velocities. The $ \gamma $-ray escape time of a homogeneous ball of iron with outer velocity $v_{\rm edge}$ and mass $M$ is
\begin{equation}
t_0=85{\rm d}~\left( \frac{M}{1.5~M_{\odot}} \right)^{0.5} \left( \frac{v_{\rm edge}}{5\times10^3 \kms} \right)^{-1},
\end{equation}
using an effective gamma-ray opacity of $\kappa_{\rm gamma}=0.025~\rm cm^{2}~g^{-1}$ \citep{1995ApJ...446..766S, 1999astro.ph..7015J}.
This implies that the produced iron has to be distributed to velocities extending beyond $\sim 5\times10^3 \kms$. Moreover, there is no room for significant additional mass at velocities $\lesssim 5\times10^3 \kms$. Assuming that the hydrogen carries most of the energy it is useful to write the constraint for additional mass at higher velocities in terms of the total energy. The gamma-ray escape time from the center of a homogeneous ball of hydrogen with kinetic energy $E$ and mass $M$ is,
\begin{equation}
t_0=87{\rm d}\left(\frac{M}{2~M_{\odot}}\right)\left(\frac{E}{10^{51}\rm erg}\right)^{-1/2}.
\end{equation}
For example, a Ni mass of $1.5 \msun$ extending to velocity $7\times10^3 \kms$ embedded in a hydrogen ball of $1\msun $ extending to $16\times10^3 \kms$ would have a total gamma-ray escape time of 70 days and total energy of $2\times 10^{51}~\rm erg$, which are not inconceivable. 

To get more insight into the envelope properties, we attempted simple semi-analytical model described by \cite{1980ApJ...237..541A} and \cite{1989ApJ...340..396A}. The formulation and implementation of this model are discussed in \cite{2015ApJ...806..160B} and references therein. This is a single component envelope model with radioactive \nickel\ confined at the center. This model is not fully applicable for \sn, as we observe no clear photospheric phase. However, by simply applying the model, we obtain an estimated envelope mass of $\sim0.8-2.0 $ \msun\ with a thermal energy of  $ \approx 1.5\times10^{51}$~erg.
It is to be emphasized that these parameters are the upper limits, as with the increase of these values the model light curves show more sustained and prominent photospheric phases. For these parameters, the ejecta becomes optically transparent at $ \sim45 $ days, which coincides  with the break seen in \textit{B}-band light curve, after which the light curve is solely powered by the $ \nickel\rightarrow\cobalt\rightarrow\iron $ decay chain.  This low ejecta mass is consistent with the inefficient $ \gamma $-ray trapping required for the radioactive decay model.
Although we see the presence of H, O and Ca lines in the late spectra, as we see in typical SNe II with large ejecta masses, these lines are much weaker than typical SNe II (see Fig.~\ref{fig:sp.comp269}), also suggestive of a low ejecta mass.

The large \nickel\ mass is consistent with the correlation between $ M_{Ni} $ and \textit{V}-band luminosity, as shown in Fig.~\ref{fig:nicomp}. 
In this scenario, \sn\ might have produced a very large amount of radioactive \nickel\ with a significantly stripped envelope. However, it is not clear which mechanism could produce such a large mass of \nickel. One possible  scenario is a thermonuclear explosion in a core-collapse supernovae \citep{CITE1, CITE2}, which would produce a wide distribution of \nickel\,masses \citep{CITE3}.  Pair instability explosions can also produce a large amount of \nickel\ with a wide range of luminosities, but they are also predicted to produce extended light curves \citep{2011ApJ...734..102K}, different from that observed for \sn. 

The possible production of a large amount of \nickel\ ($\sim 0.6 \msun$) in luminous and fast-declining SN II-L  SN~1979C was raised by \citet{1985AJ.....90.2303D, 1987txra.symp..402W, 1989ApJ...342L..79Y}. Later, \citet{1993A&A...273..106B} showed that the high luminosity of SN~1979C in its photospheric phase could also be explained by the reprocessing of UV light into optical by an extended low mass hydrogen-helium envelope. However, despite the similarities in luminosity between \sn\ and SN~1979C in its early phases, there is a clear break in the light-curve slope of SN~1979C near 70 days indicating a change in the dominant energy source. As the ejecta becomes optically thin, the UV reprocessing ceases. While \sn\ does not show such a break, making it challenging to explain the prolonged, ``perfectly'' linear light curve of \sn\ solely with reprocessing of UV energy.

{Another factor that may affect the \nickel\, mass estimate is the asymmetry of the ejecta. \citet{1999ApJ...521..179H} showed that for Type Ic-BL supernova 1998bw, an ejecta axis ratio of 2 may produce a 2 magnitudes of variation in  peak luminosity between polar and equatorial directions. Polarimetric observations of SN~1998bw were also consistent with asymmetry in the ejecta of 1998bw. Polarimetry studies suggest that CCSNe, particularly those with relatively small envelope masses, may exhibit significant asymmetry in their ejecta \citep[see e.g.][]{2001ApJ...550.1030W}. If the ejecta of \sn\, were highly asymmetric and viewed at a favorable angle, the actual amount of synthesized \nickel\ might be substantially reduced from the current estimate and shall produce its high peak luminosity. On the other hand, there is no strong evidence of asymmetry in the [\Oi] (\ldld6300, 6364) nebular lines as suggested by \cite{2009MNRAS.397..677T,2013ApJ...775L..43T}. However, quantitative modeling is needed to investigate whether an asymmetric model could explain the prolonged linear light-curve and spectroscopic features of \sn.}
 
Synthesizing a large amount of radioactive \nickel\ normally should also lead to nebular-phase spectra rich with iron-group elements, as seen in Type Ia SNe, but this doesn't appear to be the case with \sn. This could pose a serious challenge to the high \nickel\ mass scenario, and nebular-phase spectral modeling would be needed to examine it further.

\subsection{CSM interaction}\label{sec:csm}
An alternative to the radioactive decay model is strong CSM interactions. 
In these models, the structure and distribution of the CSM can determine the shape of the light curves. As in SNe~IIn \citep{1990MNRAS.244..269S}, which are characterized by narrow emission lines (FWHM $ \sim $ few hundreds of \kms) in the spectra, strong interactions are thought to be responsible for powering their prolonged and often luminous light curves \citep[e.g.][]{2006tf,2003ma}. Sometimes ejecta-CSM interactions can lead to steeply declining light curves, as has been proposed for SN 2009jp \citep[jet-CSM interaction;][]{2012MNRAS.420.1135S} and ASASSN-15no \citep{2018MNRAS.476..261B}. PTF10iam \citep{2016ApJ...819...35A}, which has early light curve features similar to \sn\ may also be explained by interaction .

The density profile of the CSM can be sculpted to produce the observed light curve of \sn. 
In \sn\ we do not see narrow or intermediate width emission lines from the expanding shock driving photo-ionization of the remaining CSM. It is possible to miss such emission lines if those were only present before our spectroscopic observations began at 53.0 days. However, this seems unlikely, as the continuous and linear decline of the light curve indicates a single dominant powering source throughout the evolution, which should produce similar narrow emission lines at later phases if they were present earlier. On the other hand, sometimes CSM signatures may also remain hidden if there is asymmetry or the CSM has a velocity of several thousands of \kms.

Some of the spectral
features in \sn\ can be attributed to indirect signatures of CSM interaction. 
\synow\ model could not reproduce the broad and extended \hb\ absorption feature with a single, regular single \Hi\ component, and the steadily increasing \hb\ absorption velocity may indicate the presence of a blended high-velocity absorption component. In the case of CSM interactions relatively weaker than SNe IIn, the enhanced excitation of outer ejecta can produce high-velocity \Hi\ absorption features (e.g., SNe 1999em, 2004dj \citep{2007ApJ...662.1136C}, 2009bw \citep{2012MNRAS.422.1122I}), without producing narrow emission lines. 
In some other cases, the interaction signature can remain blended with regular \Hi\ P-Cygni profiles, e.g. SNe 2012aw \citep{2013MNRAS.433.1871B}, 2013ej \citep{2015ApJ...806..160B}, where the two components can not be resolved individually, but resulting into broadening of \ha\ and \hb\ absorption profiles. In \sn, as the strength of the regular component decays, the effective position of the blended absorption trough would shift to the blue. However, this would also require a similar high velocity absorption feature in the \ha\ profile, which does not seem to be present, instead the entire absorption feature is missing in the SN. Another possibility is that that peculiar \ha\ emission line profile might be composed of multiple components arising from CSM interactions. 

The unusual rise in temperature from 50 to 135 days, which also coincided with the period of steady increase in the \hb\ velocity, might also be suggestive of CSM interaction. The increase in temperature could be due to the additional energy input from the ejecta interacting with CSM. The unusual break in the spectral continuum near 5500\AA\ may be also explained with CSM interaction. Similar enhancements of the blue continuum  have been seen in some strongly interacting SNe, such as SNe~2011hw \citep{2012MNRAS.426.1905S}, 2006jc \citep{2008ApJ...680..568S} and 2005ip \citep{2009ApJ...695.1334S}. The blue excess could be due to the blending of large number of broad and intermediate-width lines produced by the CSM interaction \citep{2012MNRAS.426.1905S,2009MNRAS.400..866C}. In the case of SN~2005ip, the CSM lines were narrow enough to be seen forming the blue excess.

Another possible signature indicative of CSM interaction is the A$ _1 $ and A$_2$ pair of features, discussed in \S\ref{sec:sp.evolution} (see Figs.~\ref{fig:sp.line}, \ref{fig:sp.comp124}). The favorable aspect of this pair of features is that, both are located at the same blue-shifted velocity of $ \sim3.6 $ \kkms, implying a common origin. This may indicate ejecta interacting with a clump or a shell of material is producing these \Hi\ features. However, it may be difficult to explain interaction powered light curve showing only these weak spectral signatures.

\subsection{Rate for \sn\ like SNe}\label{sec:rate}

We can make two simple estimates of the rate of ASASSN-15nx-like transients, one
based on a crude model of the ASAS-SN survey, and the other using a simple
scaling based on the number of Type~Ia SNe in ASAS-SN.  Roughly speaking, ASAS-SN
detects $V<17$~mag transients, which means that ASASSN-15nx could be detected
to a comoving distance of $\sim214$~Mpc, corresponding to a volume
of $V=0.041$~Gpc$^3$.  The rate implied by finding one such event is
\begin{equation}
r = 25 \left( { 4 \pi \over \Omega } \right)
\left( { \hbox{years} \over t_{eff}} \right)
\end{equation}
where $\Omega \simeq 2 \pi$ is the fraction of the sky being
surveyed for SNe at any given time and $t_{eff}$ is the
effective survey time.  Roughly speaking, ASAS-SN has been
running for 2.7~years, but finds only 70\% of the $V<17$~mag
SNe visible in its survey area, so $t_{eff} \simeq 1.8$~years.
Combining these factors gives a rough rate estimate of
$r \simeq 28$~Gpc$^{-3}$~year$^{-1}$. Alternatively, ASASSN-15nx
is about one magnitude more luminous than a typical Type~Ia
SNe, implying an effective survey volume to find ASASSN-15nx-like
events that is four times greater than for Type~Ia SNe.
In its first 2.7~years, ASAS-SN found
a total of 449 Type~Ia SNe, so the ASASSN-15nx-like event rate
should be $r = r_{\rm Ia}/4/449$
correcting for the larger volume for detecting ASASSN-15nx ($4$)
and the ratio of the numbers of the two events ($1:449$).  The
Type~Ia SNe rate is $r_{\rm Ia} \simeq 3 \times 10^4$~Gpc$^{-3}$~year$^{-1}$ \citep{2011MNRAS.412.1473L},
so $r \simeq 17$~Gpc$^{-3}$~year$^{-1}$, which is in good
agreement with the first estimate. Thus, the rate
of ASASSN-15nx-like transients is comparable to that for
hydrogen-poor (Type~I) SLSNe $r_{\rm SLSN-I} = 91^{+76}_{-36}$~Gpc$^{-3}$~year$^{-1}$ derived by \citet{2017MNRAS.464.3568P}, raising the possibility that the previously identified luminosity ``gap'' between normal CCSNe and SLSNe might be due to observational selection effects.

\acknowledgments
We thank A. Gal-Yam and M. Fraser for helpful comments. We are grateful to I. Arcavi for providing us the spectroscopic data for PTF10iam.
S.B., S.D., and P.C. acknowledge Project 11573003 supported by NSFC. S.B. is partially supported by China postdoctoral science foundation grant No. 2016M600848. A.P., L.T., S.B., and N.E.R. are partially supported by the PRIN-INAF 2014 project ``Transient Universe: unveiling new types of stellar explosions with PESSTO". C.S.K., K.Z.S., and T.A.T. are supported by US National Science Foundation (NSF) grants AST-1515927 and AST-1515876. We acknowledge support by the Ministry of Economy, Development, and Tourism's Millennium Science Initiative through grant IC120009, awarded to The Millennium Institute of Astrophysics, MAS, Chile (J.L.P., C.R.-C.) and from CONICYT through FONDECYT grants 3150238 (C.R.-C.) and 1151445 (J.L.P.). Support for NS was provided by the NSF through grants AST-1312221 and AST-1515559 to the University of Arizona. A.M.G. acknowledges financial support by the University of C\'adiz grant PR2017-64.

This work is based on
observations made with the Large Binocular Telescope. The
LBT is an international collaboration among institutions in
the United States, Italy, and Germany. The LBT Corporation
partners are: the University of Arizona on behalf of
the Arizona university system; the Istituto Nazionale di Astro
sica, Italy; the LBT Beteiligungsgesellschaft, Germany,
representing the Max Planck Society, theAstrophysical Institute
Potsdam, and Heidelberg University; the Ohio State
University; and the Research Corporation, on behalf of the
University of Notre Dame, the University of Minnesota, and
the University of Virginia.
We thank the staffs at the MMT Observatory for
their assistance with the observations. Observations using Steward
Observatory facilities were obtained as part of the large observing
programme AZTEC: Arizona Transient Exploration and Characterization.
Some of the observations reported in this paper were
obtained at the MMT Observatory, a joint facility of the University
of Arizona and the Smithsonian Institution.
Partially  based on observations collected at Copernico telescope (Asiago, Italy) of the INAF - Osservatorio Astronomico di Padova.

This research was made possible through the use of the AAVSO Photometric All-Sky Survey (APASS), funded by the Robert Martin Ayers Sciences Fund.
This research uses data obtained through the Telescope Access Program (TAP), which has been funded by ``the Strategic Priority Research Program-The Emergence of Cosmological Structures" of the Chinese Academy of Sciences (Grant No.11 XDB09000000) and the Special Fund for Astronomy from the Ministry of Finance.

We thank the Las Cumbres Observatory and its staff for its continuing support of the ASAS-SN project. We are grateful to M. Hardesty of the OSU ASC technology group. ASAS-SN is supported by the Gordon and Betty Moore Foundation through grant GBMF5490 to the Ohio State University and NSF grant AST-1515927. Development of ASAS-SN has  been supported by NSF grant AST-0908816, the Mt. Cuba Astronomical Foundation, the Center for Cosmology and AstroParticle Physics at the Ohio State University, the Chinese Academy of Sciences South America Center for Astronomy (CAS- SACA), the Villum Foundation, and George Skestos. This paper uses data products produced by the OIR Telescope Data Center, supported by the Smithsonian Astrophysical Observatory. 

The Liverpool Telescope is operated on the island of La Palma by Liverpool John Moores University in the Spanish Observatorio del Roque de los Muchachos of the Instituto de Astrofisica de Canarias with financial support from the UK Science and Technology Facilities Council (STFC). This paper used data obtained with the MODS spectrographs built with
funding from NSF grant AST-9987045 and the NSF Telescope System
Instrumentation Program (TSIP), with additional funds from the Ohio
Board of Regents and the Ohio State University Office of Research. ModsIDL spectral data reduction reduction pipeline developed in part with funds provided by NSF Grant AST-1108693. Partially  based on observations collected at Copernico telescope (Asiago, Italy) of the INAF - Osservatorio Astronomico di Padova. The Joan Or\'o Telescope (TJO) of the Montsec Astronomical Observatory (OADM) is owned by the Catalan Government and operated by the Institute for Space Studies of Catalonia (IEEC).

\software{MATLAB, Python, IDL, \synow\ \citep{1997ApJ...481L..89F,1999MNRAS.304...67F,2002ApJ...566.1005B}, Astropy \citep{2013A&A...558A..33A},  IRAF \citep{1993ASPC...52..173T}, LT pipeline \citep{2012AN....333..101B,2014SPIE.9147E..8HP}, DAOPHOT \citep{1987PASP...99..191S}, \textsc{FOSCGUI}, modsIDL pipeline} 
 
\bibliographystyle{apj} 
\bibliography{ms}

\clearpage
% tab:photsn
%____________________________________________________________________________

\begin{center}
	
	%\fontsize{2.0mm}{2.6mm}\selectfont%

      \tablecaption{Photometry of \sn.\label{tab:photsn_simple}}

      \tabletypesize{\fontsize{2.8mm}{3.0mm}\selectfont}

{ \startlongtable
  \begin{deluxetable*}
  	{c c r c c c c c c c l}	
  	%\caption{Photometric evolution of \sn. Errors denote $1\sigma$ uncertainty.}
  	%\label{tab:photsn}\\
  	\tablecaption{Photometry of \sn.\label{tab:photsn_simple}}
  	
  	\tabletypesize{\fontsize{1.8mm}{2.0mm}\selectfont}
  	
\tablehead{
      UT Date &         JD $-$& Phase$^a$ &                  B &                  V &                  R &                  I &                  g &                  r &                  i & Telescope$^b$ \\
              &  2,457,000    &    (days) &              (mag) &              (mag) &              (mag) &              (mag) &              (mag) &              (mag) &              (mag) &       / Inst. }
\startdata
2015-07-16.42 & 219.92 &      0.80 &                --- & >16.910  &                --- &                --- &                --- &                --- &                --- &        ASASSN \\
2015-07-19.42 & 222.92 &      3.71 &                --- & 16.540 $\pm$ 0.100 &                --- &                --- &                --- &                --- &                --- &        ASASSN \\
2015-07-21.41 & 224.91 &      5.65 &                --- & 16.370 $\pm$ 0.070 &                --- &                --- &                --- &                --- &                --- &        ASASSN \\
2015-07-23.40 & 226.90 &      7.59 &                --- & 16.200 $\pm$ 0.080 &                --- &                --- &                --- &                --- &                --- &        ASASSN \\
2015-08-08.63 & 243.13 &     23.37 &                --- & 15.830 $\pm$ 0.130 &                --- &                --- &                --- &                --- &                --- &        ASASSN \\
2015-08-09.62 & 244.12 &     24.33 &                --- & 16.000 $\pm$ 0.060 &                --- &                --- &                --- &                --- &                --- &        ASASSN \\
2015-08-09.73 & 244.23 &     24.44 & 16.183 $\pm$ 0.047 & 15.981 $\pm$ 0.016 &                --- & 15.369 $\pm$ 0.022 &                --- &                --- &                --- &         LCO \\
2015-08-12.47 & 246.97 &     27.10 &                --- & 16.010 $\pm$ 0.046 &                --- & 15.375 $\pm$ 0.039 &                --- &                --- &                --- &         LCO \\
2015-08-13.60 & 248.10 &     28.20 &                --- & 15.940 $\pm$ 0.060 &                --- &                --- &                --- &                --- &                --- &        ASASSN \\
2015-08-15.16 & 249.66 &     29.72 & 16.344 $\pm$ 0.042 & 16.037 $\pm$ 0.019 &                --- & 15.380 $\pm$ 0.026 &                --- &                --- &                --- &         LCO \\
2015-08-17.32 & 251.82 &     31.82 &                --- & 16.120 $\pm$ 0.060 &                --- &                --- &                --- &                --- &                --- &        ASASSN \\
2015-08-18.15 & 252.65 &     32.63 & 16.504 $\pm$ 0.059 & 16.132 $\pm$ 0.023 &                --- & 15.466 $\pm$ 0.025 &                --- &                --- &                --- &         LCO \\
2015-08-21.32 & 255.82 &     35.71 &                --- &                --- &                --- & 15.575 $\pm$ 0.036 &                --- &                --- &                --- &         LCO \\
2015-08-23.31 & 257.81 &     37.65 &                --- & 16.340 $\pm$ 0.080 &                --- &                --- &                --- &                --- &                --- &        ASASSN \\
2015-08-24.32 & 258.82 &     38.63 & 16.888 $\pm$ 0.058 & 16.357 $\pm$ 0.027 &                --- & 15.611 $\pm$ 0.025 &                --- &                --- &                --- &         LCO \\
2015-08-27.38 & 261.88 &     41.60 & 17.025 $\pm$ 0.070 & 16.443 $\pm$ 0.015 &                --- & 15.665 $\pm$ 0.028 &                --- &                --- &                --- &         LCO \\
2015-08-30.29 & 264.79 &     44.43 &                --- &                --- &                --- & 15.680 $\pm$ 0.033 &                --- &                --- &                --- &         LCO \\
2015-09-02.28 & 267.78 &     47.34 & 17.402 $\pm$ 0.078 & 16.637 $\pm$ 0.021 &                --- & 15.817 $\pm$ 0.036 &                --- &                --- &                --- &         LCO \\
2015-09-04.66 & 270.16 &     49.66 & 17.405 $\pm$ 0.080 & 16.625 $\pm$ 0.031 &                --- & 15.889 $\pm$ 0.027 &                --- &                --- &                --- &         LCO \\
2015-09-11.33 & 276.83 &     56.15 & 17.580 $\pm$ 0.157 & 16.923 $\pm$ 0.018 &                --- &                --- &                --- &                --- &                --- &         LCO \\
2015-09-17.33 & 282.83 &     61.98 & 17.579 $\pm$ 0.055 & 17.055 $\pm$ 0.022 &                --- & 16.212 $\pm$ 0.020 &                --- &                --- &                --- &         LCO \\
2015-09-22.61 & 288.11 &     67.12 &                --- &                --- &                --- & 16.247 $\pm$ 0.132 &                --- &                --- &                --- &         LCO \\
2015-09-26.23 & 291.73 &     70.63 & 17.974 $\pm$ 0.106 & 17.233 $\pm$ 0.031 &                --- & 16.540 $\pm$ 0.041 &                --- &                --- &                --- &         LCO \\
2015-09-28.60 & 294.10 &     72.94 & 17.870 $\pm$ 0.129 & 17.302 $\pm$ 0.056 &                --- & 16.560 $\pm$ 0.041 &                --- &                --- &                --- &         LCO \\
2015-10-02.42 & 297.92 &     76.66 &                --- & 17.401 $\pm$ 0.020 & 17.250 $\pm$ 0.014 & 16.736 $\pm$ 0.014 &                --- &                --- &                --- &         LOTIS \\
2015-10-03.42 & 298.92 &     77.63 &                --- & 17.496 $\pm$ 0.018 & 17.350 $\pm$ 0.009 & 16.674 $\pm$ 0.014 &                --- &                --- &                --- &         LOTIS \\
2015-10-11.59 & 307.09 &     85.57 & 18.125 $\pm$ 0.083 & 17.559 $\pm$ 0.036 &                --- & 16.851 $\pm$ 0.044 &                --- &                --- &                --- &         LCO \\
2015-10-12.13 & 307.63 &     86.10 &                --- &                --- &                --- &                --- & 17.829 $\pm$ 0.005 & 17.551 $\pm$ 0.008 & 17.464 $\pm$ 0.008 &           NOT \\
2015-10-15.71 & 311.21 &     89.58 & 18.201 $\pm$ 0.071 & 17.757 $\pm$ 0.035 &                --- & 17.002 $\pm$ 0.042 &                --- &                --- &                --- &         LCO \\
2015-10-21.05 & 316.55 &     94.78 & 18.261 $\pm$ 0.080 & 17.872 $\pm$ 0.034 &                --- & 17.072 $\pm$ 0.058 &                --- &                --- &                --- &         LCO \\
2015-10-23.07 & 318.57 &     96.74 & 18.348 $\pm$ 0.015 & 17.823 $\pm$ 0.015 & 17.674 $\pm$ 0.022 & 17.140 $\pm$ 0.026 & 17.948 $\pm$ 0.015 & 17.644 $\pm$ 0.021 & 17.743 $\pm$ 0.029 &     Coper,TJO \\
2015-10-25.88 & 321.38 &     99.47 &                --- &                --- &                --- &                --- &                --- &                --- &                --- &         LCO \\
2015-10-29.40 & 324.90 &    102.89 &                --- & 18.017 $\pm$ 0.070 &                --- &                --- &                --- & 17.927 $\pm$ 0.070 & 17.961 $\pm$ 0.076 &         LCO \\
2015-11-01.10 & 327.60 &    105.52 & 18.459 $\pm$ 0.052 & 17.880 $\pm$ 0.035 & 17.831 $\pm$ 0.037 & 17.371 $\pm$ 0.056 &                --- &                --- &                --- &           TJO \\
2015-11-02.86 & 329.36 &    107.23 & 18.478 $\pm$ 0.071 & 18.045 $\pm$ 0.033 &                --- &                --- &                --- & 17.909 $\pm$ 0.027 & 18.003 $\pm$ 0.046 &         LCO \\
2015-11-05.17 & 331.67 &    109.48 & 18.635 $\pm$ 0.024 & 18.216 $\pm$ 0.016 &                --- &                --- & 18.326 $\pm$ 0.014 & 17.997 $\pm$ 0.013 & 18.130 $\pm$ 0.023 &         Coper \\
2015-11-07.08 & 333.58 &    111.34 & 18.690 $\pm$ 0.023 & 18.092 $\pm$ 0.032 & 17.980 $\pm$ 0.029 & 17.496 $\pm$ 0.038 &                --- &                --- &                --- &           TJO \\
2015-11-07.32 & 333.82 &    111.57 &                --- & 18.303 $\pm$ 0.036 & 18.085 $\pm$ 0.021 & 17.741 $\pm$ 0.047 &                --- &                --- &                --- &         LOTIS \\
2015-11-07.35 & 333.85 &    111.60 & 18.684 $\pm$ 0.077 & 18.262 $\pm$ 0.035 &                --- &                --- &                --- & 18.038 $\pm$ 0.026 & 18.110 $\pm$ 0.048 &         LCO \\
2015-11-08.07 & 334.57 &    112.30 & 18.725 $\pm$ 0.020 & 18.076 $\pm$ 0.023 & 17.897 $\pm$ 0.054 & 17.506 $\pm$ 0.035 &                --- &                --- &                --- &           TJO \\
2015-11-08.32 & 334.82 &    112.54 &                --- & 18.384 $\pm$ 0.036 & 18.144 $\pm$ 0.030 & 17.593 $\pm$ 0.028 &                --- &                --- &                --- &         LOTIS \\
2015-11-09.05 & 335.55 &    113.25 & 18.742 $\pm$ 0.023 & 18.139 $\pm$ 0.024 & 18.023 $\pm$ 0.033 & 17.537 $\pm$ 0.029 &                --- &                --- &                --- &           TJO \\
2015-11-09.32 & 335.82 &    113.52 &                --- & 18.329 $\pm$ 0.031 & 18.165 $\pm$ 0.021 & 17.724 $\pm$ 0.031 &                --- &                --- &                --- &         LOTIS \\
2015-11-10.07 & 336.57 &    114.25 & 18.730 $\pm$ 0.042 & 18.144 $\pm$ 0.050 & 18.013 $\pm$ 0.035 & 17.508 $\pm$ 0.040 &                --- &                --- &                --- &           TJO \\
2015-11-11.32 & 337.82 &    115.46 &                --- & 18.353 $\pm$ 0.052 & 18.164 $\pm$ 0.025 & 17.630 $\pm$ 0.035 &                --- &                --- &                --- &         LOTIS \\
2015-11-12.07 & 338.57 &    116.19 & 18.806 $\pm$ 0.022 & 18.203 $\pm$ 0.025 & 18.087 $\pm$ 0.027 & 17.683 $\pm$ 0.036 &                --- &                --- &                --- &           TJO \\
2015-11-13.05 & 339.55 &    117.15 & 18.839 $\pm$ 0.023 & 18.253 $\pm$ 0.024 & 18.134 $\pm$ 0.024 & 17.622 $\pm$ 0.039 &                --- &                --- &                --- &           TJO \\
2015-11-13.30 & 339.80 &    117.39 & 18.754 $\pm$ 0.184 & 18.334 $\pm$ 0.033 &                --- &                --- &                --- &                --- & 18.373 $\pm$ 0.047 &         LCO \\
2015-11-14.32 & 340.82 &    118.38 &                --- & 18.420 $\pm$ 0.031 & 18.209 $\pm$ 0.026 & 17.758 $\pm$ 0.032 &                --- &                --- &                --- &         LOTIS \\
2015-11-17.04 & 343.54 &    121.02 &                --- &                --- &                --- &                --- & 18.568 $\pm$ 0.039 & 18.218 $\pm$ 0.018 & 18.422 $\pm$ 0.027 &         Coper \\
2015-11-17.32 & 343.82 &    121.30 &                --- & 18.440 $\pm$ 0.053 & 18.256 $\pm$ 0.031 & 17.684 $\pm$ 0.060 &                --- &                --- &                --- &         LOTIS \\
2015-11-19.01 & 345.51 &    122.94 & 18.876 $\pm$ 0.023 & 18.491 $\pm$ 0.019 &                --- &                --- & 18.574 $\pm$ 0.016 & 18.251 $\pm$ 0.017 & 18.457 $\pm$ 0.019 &         Coper \\
2015-11-20.32 & 346.82 &    124.21 &                --- & 18.588 $\pm$ 0.022 & 18.361 $\pm$ 0.016 & 17.929 $\pm$ 0.024 &                --- &                --- &                --- &         LOTIS \\
2015-11-22.17 & 348.67 &    126.01 & 18.949 $\pm$ 0.107 & 18.599 $\pm$ 0.038 &                --- &                --- &                --- & 18.411 $\pm$ 0.028 & 18.560 $\pm$ 0.053 &         LCO \\
2015-11-23.32 & 349.82 &    127.13 &                --- &                --- &                --- & 17.936 $\pm$ 0.573 &                --- &                --- &                --- &         LOTIS \\
2015-11-27.32 & 353.82 &    131.02 &                --- & 18.687 $\pm$ 0.097 & 18.815 $\pm$ 0.060 & 18.033 $\pm$ 0.047 &                --- &                --- &                --- &         LOTIS \\
2015-11-28.02 & 354.52 &    131.70 &                --- &                --- &                --- &                --- &                --- &                --- &                --- &         LCO \\
2015-11-29.32 & 355.82 &    132.97 &                --- & 18.825 $\pm$ 0.047 & 18.681 $\pm$ 0.040 & 18.049 $\pm$ 0.044 &                --- &                --- &                --- &         LOTIS \\
2015-11-29.52 & 356.02 &    133.16 & 19.230 $\pm$ 0.093 & 18.739 $\pm$ 0.037 &                --- &                --- &                --- & 18.538 $\pm$ 0.039 & 18.883 $\pm$ 0.086 &         LCO \\
2015-11-30.51 & 357.01 &    134.13 & 19.128 $\pm$ 0.074 & 18.771 $\pm$ 0.034 &                --- &                --- &                --- & 18.570 $\pm$ 0.035 & 18.843 $\pm$ 0.057 &         LCO \\
2015-12-03.02 & 359.52 &    136.56 & 19.207 $\pm$ 0.042 & 18.833 $\pm$ 0.026 &                --- &                --- & 19.082 $\pm$ 0.034 & 18.620 $\pm$ 0.016 & 18.882 $\pm$ 0.024 &         Coper \\
2015-12-03.33 & 359.83 &    136.87 &                --- & 18.951 $\pm$ 0.045 & 18.695 $\pm$ 0.031 & 18.270 $\pm$ 0.041 &                --- &                --- &                --- &         LOTIS \\
2015-12-03.63 & 360.13 &    137.16 &                --- & 18.713 $\pm$ 0.063 &                --- &                --- &                --- &                --- &                --- &         LCO \\
2015-12-06.84 & 363.34 &    140.28 & 19.211 $\pm$ 0.100 & 18.886 $\pm$ 0.041 &                --- &                --- &                --- & 18.707 $\pm$ 0.037 & 18.852 $\pm$ 0.081 &         LCO \\
2015-12-07.33 & 363.83 &    140.76 &                --- & 18.778 $\pm$ 0.105 & 18.790 $\pm$ 0.103 &                --- &                --- &                --- &                --- &         LOTIS \\
2015-12-09.64 & 366.14 &    143.00 &                --- & 18.759 $\pm$ 0.104 &                --- &                --- &                --- &                --- &                --- &         LCO \\
2015-12-10.33 & 366.83 &    143.67 &                --- & 19.002 $\pm$ 0.051 & 18.877 $\pm$ 0.047 & 18.277 $\pm$ 0.066 &                --- &                --- &                --- &         LOTIS \\
2015-12-13.98 & 370.48 &    147.23 & 19.579 $\pm$ 0.122 & 19.106 $\pm$ 0.066 &                --- &                --- &                --- & 18.899 $\pm$ 0.056 &                --- &         LCO \\
2015-12-16.33 & 372.83 &    149.51 &                --- & 19.255 $\pm$ 0.063 & 19.029 $\pm$ 0.033 & 18.587 $\pm$ 0.099 &                --- &                --- &                --- &         LOTIS \\
2015-12-18.01 & 374.51 &    151.14 & 19.621 $\pm$ 0.042 & 19.346 $\pm$ 0.043 &                --- &                --- & 19.362 $\pm$ 0.020 & 19.008 $\pm$ 0.035 &                --- &         Coper \\
2015-12-18.28 & 374.78 &    151.40 &                --- & 19.243 $\pm$ 0.048 &                --- &                --- &                --- & 19.075 $\pm$ 0.046 &                --- &         LCO \\
2015-12-18.90 & 375.40 &    152.01 &                --- & 19.217 $\pm$ 0.024 &                --- &                --- & 19.388 $\pm$ 0.021 & 19.033 $\pm$ 0.020 & 19.378 $\pm$ 0.027 &         Coper \\
2015-12-19.33 & 375.83 &    152.43 &                --- & 19.349 $\pm$ 0.068 & 19.188 $\pm$ 0.076 &                --- &                --- &                --- &                --- &         LOTIS \\
2015-12-23.86 & 380.36 &    156.83 &                --- & 19.225 $\pm$ 0.180 &                --- &                --- &                --- & 19.253 $\pm$ 0.214 &                --- &         LCO \\
2015-12-27.88 & 384.38 &    160.74 & 19.815 $\pm$ 0.041 & 19.445 $\pm$ 0.036 &                --- &                --- &                --- & 19.198 $\pm$ 0.028 & 19.495 $\pm$ 0.039 &            LT \\
2015-12-28.21 & 384.71 &    161.06 & 20.114 $\pm$ 0.186 & 19.464 $\pm$ 0.049 &                --- &                --- &                --- & 19.189 $\pm$ 0.041 & 19.446 $\pm$ 0.084 &         LCO \\
2015-12-30.24 & 386.74 &    163.04 &                --- & 19.528 $\pm$ 0.213 &                --- &                --- &                --- & 19.177 $\pm$ 0.099 & 19.368 $\pm$ 0.107 &         LCO \\
2016-01-01.27 & 388.77 &    165.01 &                --- & 19.630 $\pm$ 0.048 & 19.440 $\pm$ 0.047 & 19.046 $\pm$ 0.080 &                --- &                --- &                --- &         LOTIS \\
2016-01-03.27 & 390.77 &    166.96 &                --- & 19.833 $\pm$ 0.146 & 19.475 $\pm$ 0.097 & 18.889 $\pm$ 0.088 &                --- &                --- &                --- &         LOTIS \\
2016-01-05.92 & 393.42 &    169.53 & 20.198 $\pm$ 0.021 & 19.573 $\pm$ 0.018 &                --- &                --- &                --- & 19.391 $\pm$ 0.016 & 19.601 $\pm$ 0.022 &            LT \\
2016-01-10.94 & 398.44 &    174.41 & 20.347 $\pm$ 0.017 & 19.719 $\pm$ 0.017 &                --- &                --- &                --- & 19.574 $\pm$ 0.013 & 19.761 $\pm$ 0.025 &            LT \\
2016-01-12.27 & 399.77 &    175.71 &                --- & 19.876 $\pm$ 0.145 &                --- &                --- &                --- &                --- &                --- &         LOTIS \\
2016-01-16.90 & 404.40 &    180.21 & 20.487 $\pm$ 0.051 & 19.842 $\pm$ 0.042 &                --- &                --- &                --- & 19.697 $\pm$ 0.028 &                --- &            LT \\
2016-01-18.20 & 405.70 &    181.48 &                --- & 19.743 $\pm$ 0.147 & 19.633 $\pm$ 0.089 &                --- &                --- &                --- &                --- &         LOTIS \\
2016-01-21.87 & 409.37 &    185.05 & 20.546 $\pm$ 0.107 &                --- &                --- &                --- &                --- &                --- &                --- &            LT \\
2016-01-26.89 & 414.39 &    189.93 & 20.783 $\pm$ 0.016 & 20.192 $\pm$ 0.018 &                --- &                --- &                --- & 20.038 $\pm$ 0.014 & 20.242 $\pm$ 0.027 &            LT \\
2016-01-27.13 & 414.63 &    190.16 &                --- & 20.187 $\pm$ 0.155 & 19.931 $\pm$ 0.076 & 19.464 $\pm$ 0.115 &                --- &                --- &                --- &         LOTIS \\
2016-01-30.13 & 417.63 &    193.08 &                --- & 20.244 $\pm$ 0.095 & 20.226 $\pm$ 0.085 & 19.748 $\pm$ 0.120 &                --- &                --- &                --- &         LOTIS \\
2016-02-05.13 & 423.63 &    198.91 &                --- & 20.320 $\pm$ 0.186 &                --- &                --- &                --- &                --- &                --- &         LOTIS \\
2016-02-08.13 & 426.63 &    201.83 &                --- & 20.478 $\pm$ 0.149 &                --- & 19.784 $\pm$ 0.213 &                --- &                --- &                --- &         LOTIS \\
2016-02-11.10 & 429.60 &    204.72 &                --- & 20.468 $\pm$ 0.117 & 20.523 $\pm$ 0.102 &                --- &                --- &                --- &                --- &         LOTIS \\
2016-02-24.13 & 442.63 &    217.39 &                --- & 20.729 $\pm$ 0.049 &                --- &                --- &                --- &                --- &                --- &           MDM \\
2016-02-25.12 & 443.62 &    218.36 &                --- & 20.625 $\pm$ 0.030 &                --- &                --- &                --- &                --- &                --- &           MDM \\
2016-03-03.85 & 451.35 &    225.87 &                --- &                --- &                --- &                --- &                --- & 20.768 $\pm$ 0.082 &                --- &            LT \\
2016-03-04.85 & 452.35 &    226.85 &                --- &                --- &                --- &                --- &                --- & 20.819 $\pm$ 0.094 &                --- &            LT \\
2016-03-09.87 & 457.37 &    231.73 &                --- & 20.924 $\pm$ 0.045 &                --- &                --- &                --- &                --- &                --- &            LT \\
2016-03-12.10 & 459.60 &    233.90 &                --- & 21.182 $\pm$ 0.022 &                --- &                --- &                --- &                --- &                --- &         Mag \\
2016-03-15.88 & 463.38 &    237.58 &                --- & 21.213 $\pm$ 0.029 &                --- &                --- & 21.598 $\pm$ 0.049 & 21.163 $\pm$ 0.033 & 20.977 $\pm$ 0.033 &        LT,NOT \\
2016-04-09.99 & 488.49 &    261.99 & 22.290 $\pm$ 0.060 & 21.709 $\pm$ 0.049 &                --- &                --- &                --- & 21.491 $\pm$ 0.028 & 21.317 $\pm$ 0.043 &         Mag \\
\enddata

\tablecomments{\\
	$^{a}$Rest-frame days with reference to the explosion epoch \EpEpoch.\\
$^{b}$ The abbreviations of telescope/instrument used are as follows: ASASSN - ASAS-SN quadruple 14-cm telescopes; LCO - Las Cumbres Observatory 1 m telescope
	network; LT - 2m Liverpool Telescope; NOT - ALFOSC mounted on 2.0m NOT telescope; MDM - 2.4m MDM telescope ; LOTIS - 0.6m Super-Lotis telescope; TJO -  0.8m TJO telescope; Coper - 1.8m Copernico telescope; Mag - IMACS mounted on 6.5m Magellan Baade telescope.\\
	%$^{c}$ FWHM of the median stellar PSF at $V$ band frame.\\
	Data observed within 5\,hr are represented under a single-epoch observation.}
\end{deluxetable*}
}

%\begin{flushleft}
%  $^{a}$ with reference to the explosion epoch \EpEpoch\\
%  $^{b}$ ST : 104-cm Sampurnanand Telescope, ARIES, India; DFOT : 130-cm Devasthal fast optical telescope, ARIES, India; HCT: 2m Himalyan Chandra Telescope, Hanle, India; Swift: \textit{Swift}~UVOT\\
%  %$^{c}$ FWHM of the median stellar PSF at $V$ band frame.\\
%  Note: Data observed within 5 Hrs, are represented under single epoch observation.
%\end{flushleft}
\end{center}

\clearpage

% tab:speclog
%____________________________________________________________________________

\begin{table}
\centering
  \caption{Summary of spectroscopic observations of \sn.}
  \label{tab:speclog}
  \begin{tabular}{lc cc cc}
    \hline
    UT Date       &JD      &Phase$^{a}$&Telescope$^{b}$ \\
    (yy/mm/dd.dd) &2457000+&(days)     &                \\ \hline
2015-09-08.12  &  273.62   &   53.0    &  Coper  \\
2015-09-08.38  &  273.88   &   53.3    &  Dup  \\
2015-10-13.46  &  308.96   &   87.4    &  MOD  \\
2015-10-22.00  &  318.50   &   96.7    &  Coper  \\
2015-10-23.50  &  319.00   &   97.2    &  Bok  \\
2015-11-05.09  &  331.59   &   109.4   &  NOT  \\
2015-11-05.10  &  331.60   &   109.4   &  Coper  \\
2015-11-16.99  &  343.49   &   121.0   &  Coper  \\
2015-11-18.98  &  345.48   &   122.9   &  Coper  \\
2015-11-19.30  &  345.80   &   123.2   &  Bok  \\
2015-12-01.32  &  357.82   &   134.9   &  MMT  \\
2015-12-02.97  &  359.47   &   136.5   &  Coper  \\
2015-12-07.25  &  363.75   &   140.7   &  MOD  \\
2015-12-17.96  &  374.46   &   151.1   &  Coper  \\
2016-01-02.21  &  389.71   &   165.9   &  MOD  \\
2016-01-02.29  &  389.79   &   166.0   &  Bok  \\
2016-01-16.08  &  403.58   &   179.4   &  MagE  \\
2016-03-12.09  &  459.59   &   233.9   &  IMAC  \\
2016-04-10.01  &  488.51   &   262.0   &  IMAC  \\
    \hline
  \end{tabular}
\begin{flushleft}
  $^{a}$ The phase is the number of rest frame days after the adopted  explosion epoch \EpEpoch\\
  $^{b}$ The telescope abbreviations are - Coper : Copernico telescope; Dup : Du Pont telescope; MOD: MODS spectrograph on LBT; Bok: Bok telescope; MMT: MMT Observatory; NOT: Nordic Optical Telescope; MagE: Echellette Spectrograph on Magellan Baade telescope; IMAC: IMACS spectrograph on Magellan Baade telescope. \\
\end{flushleft}
\end{table}

\begin{table}
\centering
  \caption{Black-body Bolometric luminosity.}
  \label{tab:BB}
  \fontsize{2.5mm}{2.7mm}\selectfont
  \begin{tabular}{r c | r c}
    \hline
Phase$^{a}$  &Luminosity                   &Phase$^{a}$  &Luminosity     \\
(days)      & ($Log_{10}[L~erg~s^{-1}]$)  & (days)      & ($Log_{10}[L~erg~s^{-1}]$) \\ \hline
3.72   & 43.49 $\pm$ 0.17  &    118.38 & 42.46 $\pm$ 0.40\\
5.65   & 43.53 $\pm$ 0.17  &    121.02 & 42.44 $\pm$ 0.13\\
7.59   & 43.56 $\pm$ 0.17  &    121.30 & 42.45 $\pm$ 0.13\\
23.37  & 43.48 $\pm$ 0.17  &    122.94 & 42.43 $\pm$ 0.11\\
24.44  & 43.50 $\pm$ 0.20  &    124.21 & 42.40 $\pm$ 0.12\\
27.10  & 43.47 $\pm$ 0.17  &    126.01 & 42.40 $\pm$ 0.89\\
28.20  & 43.45 $\pm$ 0.16  &    127.13 & 42.38 $\pm$ 0.63\\
29.72  & 43.44 $\pm$ 0.14  &    131.02 & 42.35 $\pm$ 0.24\\
31.82  & 43.38 $\pm$ 0.16  &    131.70 & 42.34 $\pm$ 0.23\\
32.63  & 43.38 $\pm$ 0.15  &    132.97 & 42.32 $\pm$ 0.25\\
35.71  & 43.32 $\pm$ 0.15  &    133.16 & 42.31 $\pm$ 0.17\\
37.65  & 43.28 $\pm$ 0.13  &    134.12 & 42.33 $\pm$ 0.17\\
38.63  & 43.27 $\pm$ 0.11  &    135.30 & 42.31 $\pm$ 0.17\\
41.61  & 43.23 $\pm$ 0.11  &    135.52 & 42.31 $\pm$ 0.14\\
44.44  & 43.21 $\pm$ 0.12  &    136.34 & 42.30 $\pm$ 0.13\\
47.34  & 43.16 $\pm$ 0.10  &    136.56 & 42.30 $\pm$ 0.13\\
49.66  & 43.15 $\pm$ 0.10  &    136.87 & 42.28 $\pm$ 0.19\\
56.14  & 43.06 $\pm$ 0.21  &    137.16 & 42.31 $\pm$ 0.14\\
61.98  & 43.00 $\pm$ 0.10  &    140.28 & 42.29 $\pm$ 0.25\\
63.02  & 42.99 $\pm$ 0.10  &    140.76 & 42.28 $\pm$ 0.21\\
64.56  & 42.97 $\pm$ 0.16  &    143.00 & 42.27 $\pm$ 0.19\\
67.12  & 42.96 $\pm$ 0.21  &    143.67 & 42.23 $\pm$ 0.19\\
70.64  & 42.90 $\pm$ 0.15  &    147.22 & 42.17 $\pm$ 0.24\\
72.94  & 42.89 $\pm$ 0.22  &    149.51 & 42.14 $\pm$ 0.18\\
76.66  & 42.84 $\pm$ 0.19  &    151.14 & 42.12 $\pm$ 0.19\\
77.63  & 42.82 $\pm$ 0.21  &    151.41 & 42.12 $\pm$ 0.16\\
85.57  & 42.78 $\pm$ 0.15  &    152.01 & 42.12 $\pm$ 0.17\\
86.10  & 42.76 $\pm$ 0.16  &    152.43 & 42.10 $\pm$ 0.17\\
89.58  & 42.72 $\pm$ 0.16  &    156.83 & 42.09 $\pm$ 0.20\\
94.77  & 42.69 $\pm$ 0.18  &    160.74 & 42.05 $\pm$ 0.16\\
96.74  & 42.68 $\pm$ 0.08  &    161.12 & 42.00 $\pm$ 0.25\\
99.47  & 42.65 $\pm$ 0.16  &    165.01 & 41.96 $\pm$ 0.37\\
102.90 & 42.62 $\pm$ 0.16  &    166.96 & 41.93 $\pm$ 0.24\\
105.52 & 42.63 $\pm$ 0.10  &    169.53 & 41.95 $\pm$ 0.12\\
107.23 & 42.60 $\pm$ 0.15  &    174.42 & 41.90 $\pm$ 0.10\\
109.48 & 42.54 $\pm$ 0.11  &    175.71 & 41.86 $\pm$ 0.16\\
111.34 & 42.55 $\pm$ 0.08  &    180.21 & 41.85 $\pm$ 0.15\\
111.57 & 42.51 $\pm$ 0.16  &    181.48 & 41.86 $\pm$ 0.15\\
111.60 & 42.52 $\pm$ 0.18  &    185.05 & 41.80 $\pm$ 0.23\\
112.30 & 42.55 $\pm$ 0.08  &    187.31 & 41.76 $\pm$ 0.20\\
112.54 & 42.50 $\pm$ 0.12  &    189.93 & 41.72 $\pm$ 0.13\\
113.25 & 42.53 $\pm$ 0.08  &    190.16 & 41.73 $\pm$ 0.16\\
113.52 & 42.49 $\pm$ 0.12  &    193.08 & 41.68 $\pm$ 0.14\\
114.24 & 42.54 $\pm$ 0.11  &    194.79 & 41.68 $\pm$ 0.13\\
115.46 & 42.48 $\pm$ 0.11  &    195.74 & 41.67 $\pm$ 0.14\\
116.19 & 42.50 $\pm$ 0.09  &    198.91 & 41.65 $\pm$ 0.22\\
117.14 & 42.49 $\pm$ 0.09  &    201.83 & 41.61 $\pm$ 0.27\\
117.39 & 42.49 $\pm$ 0.29  &    261.99 & 41.32 $\pm$ 0.26\\
    \hline
  \end{tabular}
\begin{flushleft}
	The bolometric luminosity is calculated by fitting black-body on \textit{BVI} bands photometric data. During pre-peak phases ($ \leq 23.37 $d), where only \textit{V}-band data available, the blackbody temperature is extrapolated and luminosity is calculated after scaling the blackbody SED to match \textit{V}-band data, also taking into account of the broadband filter response. At 261.99 days the fitting is done on $ BVri $-bands data. \\
	$^{a}$Rest-frame days relative to the adopted explosion epoch \EpEpoch.\\
\end{flushleft}
\end{table}

\end{document}